\documentclass[9pt,twocolumn]{iopart}

\usepackage{calrsfs}
\DeclareMathAlphabet{\pazocal}{OMS}{zplm}{m}{n}
\usepackage{amsfonts}       
\usepackage{graphicx}
\usepackage{dcolumn}
\usepackage{epsfig}
\usepackage{bm}
\usepackage{array}
\usepackage{hyperref}
\hypersetup{
colorlinks=true,
citecolor=blue,
linkcolor=blue,
urlcolor=magenta,
}
\usepackage{xcolor}			
\usepackage{graphicx}
\usepackage{ulem}
\usepackage{amssymb}
\usepackage{bm}           
\usepackage{bbm}
\usepackage[english]{babel}
\usepackage{setspace}
\usepackage{wasysym}
\usepackage{adjustbox}
\usepackage{framed}
\usepackage{mdframed}
\usepackage{latexsym} 
\usepackage{pdflscape}   
\usepackage{multirow}    
\usepackage{booktabs}    
\usepackage{calc}        
\usepackage{cite}

\definecolor{DeepSkyBlue}{RGB}{0,104,139}
\colorlet{MySky}{white!40!blue}
\colorlet{MyViolet}{red!45!blue}
\colorlet{MyBlue}{black!40!blue}
\colorlet{MyRed}{black!40!red}
\colorlet{MyOrange}{red!70!yellow}
\colorlet{MyGreen}{black!50!green}
\colorlet{MyBrown}{black!70!brown}
\colorlet{MyGray}{black!60!white}

\newcommand{\OP}{\Phi}
\newcommand{\pow}{\beta}

\newcommand{\vect}[1]{\bm{#1}}

\newcommand{\be}{\begin{equation}}
\newcommand{\ee}{\end{equation}}

\newcommand{\bra}[1]{\ensuremath{\langle#1|}}
\newcommand{\ket}[1]{\ensuremath{|#1\rangle}}
\newcommand{\mean}[1]{\ensuremath{\langle #1 \rangle}}

\newcommand{\spinupz}{|\!\!\,\uparrow_z\rangle}
\newcommand{\spindownz}{|\!\!\,\downarrow_z\rangle}
\newcommand{\spinupx}{|\!\!\,\uparrow_x\rangle}
\newcommand{\spindownx}{|\!\!\,\downarrow_x\rangle}
\newcommand{\spinupn}{|\!\!\,\uparrow_{\vect{n}}\rangle}
\newcommand{\spindownn}{|\!\!\,\downarrow_{\vect{n}}\rangle}

\newcommand{\Eunit}{\pazocal{J}}
\newcommand{\thetac}{\theta_{\rm c}}
\newcommand{\thetaFM}{\theta_{\rm c}^{\,-}}		
\newcommand{\thetaAFM}{\theta_{\rm c}^{\,+}}		

\begin{document} 

\title[Multipartite-Entanglement Tomography of a Quantum Simulator]{Multipartite-Entanglement Tomography of a Quantum Simulator}

\author{Marco Gabbrielli$^{1,*}$, Luca Lepori$^{2,3,4,*}$ and Luca Pezz\`e$^1$}
\address{$^1$ QSTAR and INO-CNR and LENS, largo Enrico Fermi 2, I-50125 Firenze, Italy} 
\address{$^2$ Dipartimento di Scienze Fisiche e Chimiche, Universit\`a degli Studi dell'Aquila, via Vetoio, I-67010 Coppito, L'Aquila, Italy} 
\address{$^3$ INFN, Laboratori Nazionali del Gran Sasso, via Giovanni Acitelli 22, I-67100 Assergi, L'Aquila, Italy} 
\address{$^4$ Istituto Italiano di Tecnologia, Graphene Labs, Via Morego 30, I-16163 Genova,~Italy}
\address{$^*$ {\it These authors contributed equally to the present work.}}
\ead{luca.pezze@ino.it}

\begin{abstract} 
Multipartite-entanglement tomography, namely 
the quantum Fisher information (QFI) calculated with respect to different collective operators, 
allows to fully characterize the phase diagram of the quantum Ising chain in a transverse field with variable-range coupling. 
In particular, it recognizes the phase stemming from long-range antiferromagnetic coupling, 
a capability also shared by the spin squeezing.
Furthermore, the QFI locates the quantum critical points, both with vanishing and nonvanishing mass gap.
In this case, we also relate the finite-size power-law exponent of the QFI to the critical exponents of the model, finding
a signal for the breakdown of conformal invariance in the deep long-range regime.
Finally, the effect of a finite temperature on the multipartite entanglement, and ultimately on the phase stability, is considered.
In light of the current realizations of the model with trapped ions  and of the potential measurability of the QFI, our approach yields a promising
strategy to probe long-range physics in controllable quantum systems. 
\end{abstract} 


\maketitle

\section{Introduction} 
The experimental realization of quantum simulators~\cite{CiracNATPHYS2012, GeorgescuRMP2014} 
has made a significant progress in the recent years~\cite{SimonNATURE2011, BrittonNATURE2012, IslamSCIENCE2013, 
RichermeNATURE2014, JurcevicNATURE2014, ZhangNATURE2017, BernienNATURE2017}:
systems of trapped ions~\cite{BlattNATPHYS2012, SchneiderRPP2012}, ultracold atoms and molecules~\cite{BlochNATPHYS2012, LewensteinBOOK, InguscioBOOK}
and superconducting circuits~\cite{DevoretSCIENCE2013} are currently able to simulate important models of quantum physics. 
A  notable example is the long-range quantum Ising chain in a transverse field, which has been realized
with up to $\sim50$ spins~\cite{ZhangNATURE2017, BernienNATURE2017}.
The experiments are rapidly approaching the point where the outcomes cannot be efficiently computed on a classical machine. 
We thus need methods for the reliable benchmarking of quantum simulators~\cite{HaukeRPP2012,MartyPRL2016}. 
These might be given, for instance, by detecting specific properties of the ground state of the system that can be accessed without full state tomography.

The measurement of a local order parameter is a standard example of such benchmarking: it signals the onset of a dominant order in the system 
when tuning a control parameter that rules the competition between non-commuting terms in a many-body Hamiltonian. 
It has thus been used to detect a variety of quantum phase transitions (QPTs), in analogy to the detection of thermal phase transitions.
This approach, however, provides no information about quantum correlations in the considered system.
Moreover, a local order parameter cannot distinguish between topologically trivial and nontrivial phases~\cite{BernevigBOOK}. 

Another approach, which has emerged in the last decades~\cite{AmicoRMP2008, EisertRMP2010, Zeng}, is to characterize the system via the 
bipartite entanglement (BE) properties of the ground state.
Entanglement between two parts of a many-body system is a pivotal figure of merit and
it is  analyzed typically via the Von Neumann entropy~\cite{AmicoRMP2008, Zeng, LatorreJPA2009, EisertRMP2010, OsbornePRA2002, VidalPRL2003} 
or  the entanglement spectrum~\cite{LiPRL2008,PollmannPRB2010,FidkowskiPRL2010,ThomalePRL2010,LeporiPRB2012}.
An alternative approach to BE is the study of the two-body reduced density matrix~\cite{OsterlohNATURE2002,LidarPRL2004}, also quoted as pairwise entanglement.
BE has attracted large attention because it can be efficiently computed~\cite{AmicoRMP2008}  
and it is a resource required for classical simulations of many-body systems with numerical methods~\cite{EisertRMP2010}.
It has been shown that in several short-range (SR) one-dimensional models BE diverges 
logarithmically with the system size at criticality, whereas it does not scale in any gapped phase~\cite{AmicoRMP2008, EisertRMP2010, Zeng}. 
Instead, for long-range (LR) models such a violation of the area law is found also in gapped phases~\cite{KoffelPRL2012, VodolaPRL2014, VodolaNJP2016, LeporiPRB2018}.
Yet, not only it is difficult to experimentally extract BE in large systems~\cite{MartyPRL2016} 
but, furthermore, a logarithmic scaling might be hard to distinguish from a constant 
behavior in systems of relatively small size.

Here we consider a further possible approach to benchmark a quantum simulator.
This is based on the susceptibility of the ground state $\ket{\psi_{\rm gs}}$ 
to unitary transformations $e^{-i \phi \hat{O}}$ generated by some operator $\hat{O}$ and parametrized by the real value $\phi$, 
as given by the quantum Fisher information (QFI)~\cite{Pezze2014, TothJPA2014, PezzeRMP}.
The QFI $F_Q[\hat{\rho}, \hat{O}]$ of a generic state $\hat{\rho}$
quantifies the ``spread'' of the state over the eigenstates of $\hat{O}$ 
(notice that $F_Q[\hat{\rho}, \hat{O}]=0$ if and only if $[\hat{O}, \hat{\rho}] = 0$)
and, in particular, it reduces to the variance $F_Q[\ket{\psi}, \hat{O}]=4 (\Delta \hat{O})^2$ for pure states.
Importantly, the QFI is a witness of multipartite entanglement (ME): for local operators $\hat{O}$, as in the case of this manuscript,
$F_Q[\hat{\rho}, \hat{O}]> N k$ detects $k$-partite entanglement among $N$ spins \cite{PezzePRL2009, HyllusPRA2012, TothPRA2012, PezzePNAS2016, GessnerPRA2016}.
In particular, ME is able to capture the richness of multiparticle correlations of many-body states beyond BE.
The QFI of a quantum states calculated with respect to different operators $\hat{O}$ provides a ``multipartite-entanglement tomography'' that gives information not only about 
ME, but also about global properties of the correlation functions. 
The QFI is thus able to recognize different phases and QPTs 
of a many-body model~\cite{MaPRA2009, LiuJPA2013, HaukeNATPHYS2016, PezzePRL2017, ZhangPRL2018, Gabbrielli2018}. 

In the present paper, we illustrate these ideas for the Ising chain with variable-range interaction in a transverse field. 
We show how multipartite-entanglement tomography based on the QFI can give information about -- and distinguish -- 
the paramagnetic, ferromagnetic and antiferromagnetic phases of the model.
For ordered phases, the optimal choice of operator $\hat{O}$ is given by the order parameter of the transition, characterized by diverging fluctuations, 
and giving a Heisenberg scaling of the QFI, $F_Q[\ket{\psi_{\rm gs}}, \hat{O}] \sim N^2$. 
For disordered phases there is an important difference between the SR and LR regimes: while in the SR case the QFI is extensive, $F_Q[\ket{\psi_{\rm gs}}, \hat{O}] \sim N$, in the LR case
the QFI is superextensive, $F_Q[\ket{\psi_{\rm gs}}, \hat{O}] \sim N^\beta$ with $1< \beta \leq 3/2$. 
This scaling law is directly related to the presence of power-law decaying correlation functions, 
where $\hat{O}$ here is a suitable collective operator -- generally different from the order parameter -- that maximizes the QFI in this regime.
Interestingly, the long-range disordered phase is also recognized by the spin-squeezing parameter.
We discuss the change of scaling of the QFI at the critical points 
when interactions change from SR to LR, suggesting the breakdown of conformal invariance and capturing the mean-field limit of the model. 
We finally extend our analysis to  finite temperature and show that the large entanglement found in the ground state of the LR disordered phase 
is robust against temperature being protected by a finite energy gap. 
Our results can be readily tested in current experimental systems. 
In particular, the finite-size power-law scaling of the QFI is thus able -- even at experimentally available sizes ($N \approx 50$) -- to detect
the appearance of long-range phases and to characterize QPTs beyond nearest-neighbor interaction. 
It is indeed worth pointing out that the QFI can be experimentally addressed: it is related to  dynamical susceptibilities \cite{HaukeNATPHYS2016},
and a lower bound can be obtained from the variation of statistical distributions of a measured observable \cite{StrobelSCIENCE2014, PezzePNAS2016}. 

\section{The model}

We study the one-dimensional quantum Ising chain in a transverse field, with variable-range coupling and open boundary conditions.
The corresponding Hamiltonian is 
\be 
\label{Hamiltonian}
\hat H = \Eunit \sin\theta \sum_{i=1}^{N-1} \sum_{j=i+1}^{N} \frac{\hat{\sigma}_z^{(i)} \hat{\sigma}_z^{(j)}}{|i-j|^\alpha} 
+ \Eunit\cos\theta\sum_{i=1}^N \hat{\sigma}_x^{(i)},
\ee
where $N$ is the number of spins (in the following we assume even $N$), 
$\hat{\sigma}_{\mathbf{n}}^{(i)}$ is the Pauli matrix for the $i$th spin ($i=1,2,...,N$) along the direction $\mathbf{n}$, 
and $\Eunit>0$ sets the energy scale. 
The parameter $\theta\in[-\pi/2,\pi/2]$ rules the competition between 
the transverse external field of magnitude $\Eunit\cos\theta$
and the spin-spin coupling of strength $\Eunit\sin\theta$.
The decay power $\alpha\geq0$ specifies the range of the spin-spin coupling, which is 
ferromagnetic (FM) for $\theta<0$ and antiferromagnetic (AFM) for $\theta>0$.
For $\alpha\to\infty$, Eq.~(\ref{Hamiltonian}) reduces to the well-known quantum Ising model with nearest-neighbor coupling \cite{PfeutyANNPHYS1970,Dutta2015}.
For $\alpha=0$, Eq.~(\ref{Hamiltonian}) corresponds to a chain with infinite-range coupling.
For finite values of $\alpha$, Eq.~(\ref{Hamiltonian}) is a paradigmatic model to study the physical effects induced by LR coupling. 
Indeed, various theoretical works pointed out that this model displays 
many interesting and peculiar features \cite{DuttaPRB2001, DengPRA2005}, ultimately connected to the 
effective violation of locality \cite{HaukePRL2013,SantosPRL2016}, including the semi-algebraic decay for correlations in gapped regimes \cite{KoffelPRL2012,VodolaNJP2016,HastingsCMP2006}, the related violation of the area law for the Von Neumann entropy \cite{KoffelPRL2012,VodolaPRL2014} and 
anomalous distribution for the entanglement spectrum \cite{KoffelPRL2012, HaukePRL2013}, and the breakdown of conformal invariance at criticality \cite{MaghrebiPRL2015,LeporiAP2016}.
Moreover, new phases displaying these features, but not belonging to the classification schemes for SR systems, have been identified theoretically in this model
\cite{KoffelPRL2012, VodolaNJP2016, GongPRB2016}.
The interesting physics associated to LR interaction concerns also fermionic lattice systems, characterized by nontrivial topological invariants
\cite{VodolaPRL2014,VodolaNJP2016,delgado2015,
gong2015, MaghrebiPRL2015, LeporiJSM2016, LeporiNJP2016, giuliano2017, LeporiPRB2018, delgado2018}.
Notably, for these systems, BE is known to characterize only partially the LR regimes, not being able to distinguish in general the different LR phases \cite{LeporiNJP2016,LeporiPRB2018}, while ME appears to be more indicative \cite{PezzePRL2017, ZhangPRL2018, Gabbrielli2018}.

Recently, the Hamiltonian (\ref{Hamiltonian}) has been 
experimentally implemented with up to $N\approx50$ spins. 
This has been performed
using trapped ions \cite{RichermeNATURE2014, IslamSCIENCE2013, JurcevicNATURE2014, ZhangNATURE2017}, 
Rydberg atoms in a cavity \cite{BernienNATURE2017}, 
and ultracold spinless atoms in an optical lattice \cite{SimonNATURE2011}.
In trapped-ion experiments, the tunable decay power $\alpha$ can be adjusted in the range $0 \lesssim \alpha \lesssim 3$. 

\section{Phase diagram}
The phase diagram of the model shown in Fig.~\ref{fig1} 
is determined by the competition between the 
two non-commuting terms in Eq.~(\ref{Hamiltonian}): 
the longitudinal exchange coupling and the transverse magnetic field.

\subsection{Critical lines}

\begin{figure}[!t]
\centering
\includegraphics[width=0.75\textwidth]{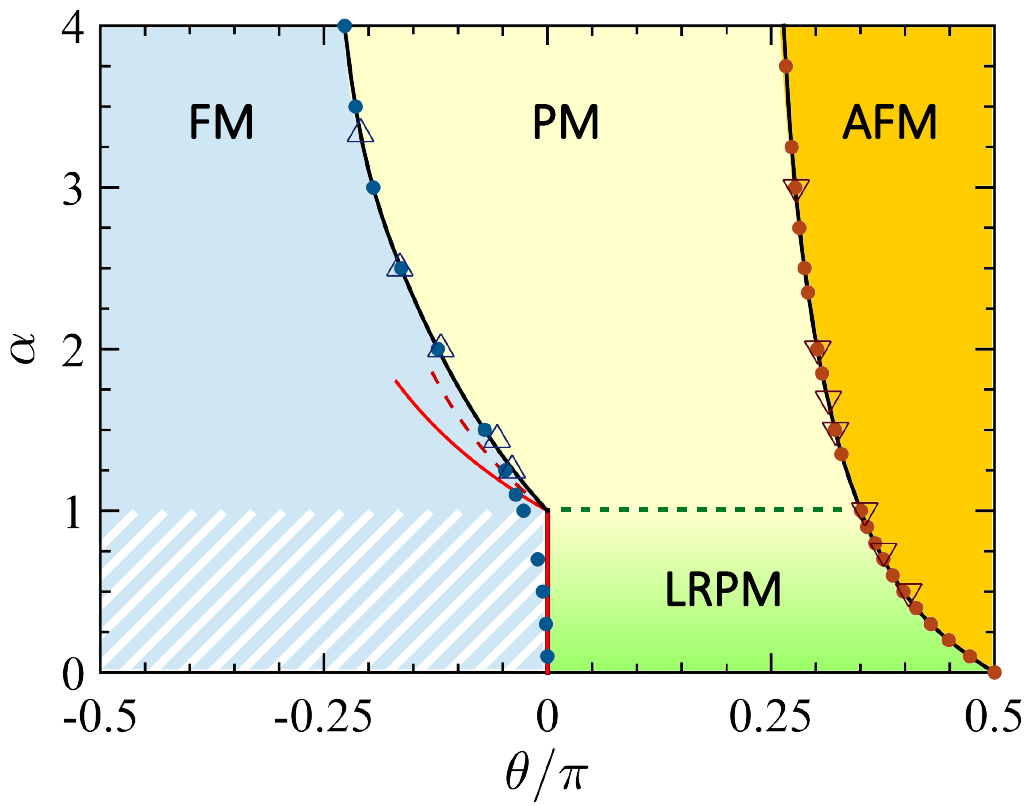} 
\caption{Phase diagram of the Ising chain in the $\theta$--$\alpha$ plane.
Colored regions highlight different phases, as recognized by both a suitable order parameter and entanglement property (see main text).
For $\alpha \leq 1$ and $\theta<0$ (hatched region), the thermodynamic limit is not defined.
The solid black lines, separating the ordered phases from the disordered one,
mark a vanishing mass gap in the thermodynamic limit, they interpolate
the numerical data $\thetaFM$ (blue dots) and $\thetaAFM$ (orange dots).
Triangles are known results in the literature for the FM transition (blue triangles, Ref.~\cite{KnapPRL2013}) 
and AFM transition (red triangles, Ref.~\cite{KoffelPRL2012}), see also Ref.~\cite{FeyPRA2016}.
The red lines show the position of the FM critical points as
calculated by a perturbative expansion at the first order (solid) and at the second order (dashed) in $\theta \to 0$.
The horizontal dashed line denotes a massive critical line at $\alpha=1$, separating the 
short-range paramagnetic (PM) phase from the long-range (LRPM) one.}
\label{fig1}
\end{figure}

For any fixed $\alpha$, the Ising chain hosts two QPTs driven by the control parameter~$\theta$.
Each QPT separates a magnetically disordered phase from an ordered one,
according to the spontaneous symmetry breaking of the spin-flip $Z_2$ invariance of the Hamiltonian (\ref{Hamiltonian}) in the thermodynamic limit. 
This behavior results in two lines of critical points $\thetaFM(\alpha)\leq0$ and $\thetaAFM(\alpha)>0$, 
where transitions from a paramagnetic (PM) phase to FM and AFM phases take place, respectively. 
For $\alpha>0$, both the critical lines signal second-order QPTs.
The model is analytically solvable in two limit cases: 
for nearest-neighbor coupling ($\alpha=\infty$) within a Jordan-Wigner transformation~\cite{Lieb1961}; and for 
infinite range coupling ($\alpha=0$), within a Bethe ansatz \cite{PanPLB1999, MoritaNPB2006} and in the thermodynamic limit~\cite{RibeiroPRL2007}.
In the case $\alpha=\infty$ the exact location of the critical points is well known \cite{Dutta2015}: 
$\thetaFM(\infty)=-\pi/4$ and $\thetaAFM(\infty)=\pi/4$.
For $\alpha=0$, instead, the fully-connected chain has a second-order FM transition at $\theta=0$ \cite{BotetPRL1982,BotetPRB1983} and a 
first-order AFM transition at $\theta=\pi/2$ \cite{VidalPRA2004}. 
For any finite value of $\alpha$, the emerging QPTs at finite $N$ are signaled by a minimum of the mass gap $\Delta_N(\alpha,\theta)$, as a function of $\theta$.
In order to locate the transitions, we determine $\theta_N^{\pm}(\alpha)=\min_\theta\Delta_N(\theta,\alpha)$ for $N=10\dots120$ and 
extrapolate the asymptotic value for $N \to \infty$ by a fit.
The numerical results are reported as dots in Fig. \ref{fig1}. 
The qualitative shape of the critical lines $\thetaFM(\alpha)$ (blue dots) and $\thetaAFM(\alpha)$ (orange dots) noticeably differ each others
as a consequence of the distinct effect of the spin-spin coupling.

For $\theta<0$, the LR coupling enforces the FM order, even at strong magnetic fields: 
at fixed $\alpha$, the PM phase progressively shrinks when increasing $N$,
and it disappears in the large-$N$ limit if $\alpha \leq 1$.
In this regime ($\theta<0$ and $\alpha \leq 1$), a perturbative calculation of the mass gap $\Delta_N$ 
at first order for small values of the control parameter $\theta$ (see \ref{AppPerturb})
provides $\theta_N^{-}(\alpha)=-1/N^{1-\alpha}$ for $\alpha<1$
and $\theta_N^{-}(\alpha)=-1/\log{N}$ for $\alpha=1$,
ensuring that $\thetaFM(\alpha)=0$ for $\alpha\leq1$ in the thermodynamic limit, 
as indicated by the red solid line in Fig.~\ref{fig1}.
We argue that the location of numerical results (blue dots) out of $\theta=0$ is a numerical artifact of the finite-size analysis.
The predictions of the perturbative calculation for $\alpha>1$ are also shown in Fig.~\ref{fig1} as red lines:
in the thermodynamic limit we predict $\thetaFM(\alpha)=-1/\zeta(\alpha)$ at first order (solid line)
and $\thetaFM(\alpha)=-(\sqrt{3}-1)/\zeta(\alpha)$ at second order (dashed line) in $\theta$. 

For $\theta>0$, instead, the LR coupling strongly frustrates the AFM order: 
frustration entails a preference for the system to endure in the  
disordered phase, even at low magnetic fields.
Consequently, the AFM critical point shifts towards larger values of $\theta$
as $\alpha$ decreases. 
In particular, the fully-connected chain $\alpha=0$ becomes completely frustrated
and the corresponding AFM phase has a vanishing extension, reducing to the single point $\thetaAFM(0)=\pi/2$.

Finally, we notice that many studies, based on different numerical methods, have investigated the 
AFM and FM critical lines~\cite{KoffelPRL2012, KnapPRL2013, FeyPRA2016, VodolaNJP2016, DefenuPRB2017, JaschkeNJP2017, SunPRA2017, ZhuARXIV2018}.
Our numerical results agree well with the literature.
In particular, in Fig.~\ref{fig1}, for comparison, we report the the location of the PM-to-FM quantum phase transition
based on scaled exact diagonalization in the FM regime (for $\alpha\gtrsim 1$)~\cite{KnapPRL2013},
and the location of the PM-to-AFM quantum phase transition based on maxima of the half-chain 
Von Neumann entropy (for $\alpha\gtrsim 0.5$)~\cite{KoffelPRL2012}.

\subsection{Quantum phases}

The characterization of the different phases bounded by $\thetaFM(\alpha)$ and $\thetaAFM(\alpha)$ 
is primarily done in terms of suitable order parameters that recognize the onset of the dominant FM and AFM order.
We can distinguish three phases, see Fig.~\ref{fig1}: \\

{\it Ferromagnetic phase --} For sufficiently strong FM interaction, $-\pi/2\leq\theta<\thetaFM (\alpha)$,
the system exhibits an ordered FM phase, where the $Z_2$ symmetry is spontaneously broken in the limit $N\to\infty$.
The order is detected by the longitudinal magnetization $\OP_z = \bra{\psi_{\mathrm{gs}}}\hat{J}_z\ket{\psi_{\mathrm{gs}}}$,
where $\hat{J}_z = \frac{1}{2} \sum_{i=1}^N \hat{\sigma}_z^{(i)}$.
$\OP_z$ is nonvanishing in a finite chain provided that  an irrelevant
$Z_2$ symmetry-breaking perturbation $h\,\hat{\sigma}_z^{(N)}$, with $h\to0$, is added to the Hamiltonian (\ref{Hamiltonian}).
If such a perturbation is not added, in the limit $\theta\to\pi/2$ the ground state is the
Greenberger-Horne-Zeilinger (GHZ) state 
$\ket{\psi_{\rm gs}} = (\spinupz^{\otimes N} 
\allowbreak + \allowbreak 
\spindownz^{\otimes N})/\sqrt{2}$
for all values of $\alpha$, while in the $N \to \infty$ limit this state becomes degenerate with $\ket{\psi^{\prime}_{\mathrm{gs}}} = (\spinupz^{\otimes N}
\allowbreak - \allowbreak 
\spindownz^{\otimes N})/\sqrt{2}$.
Here and in the following, $\spinupn$ and $\spindownn$ denote the eigenstates of $\hat{\sigma}_{\vect{n}}$.

The FM phase for $\alpha \leq 1$ (hatched region in Fig.~\ref{fig1}) deserves a comment since, here, the energy is superextensive. 
In this case, the thermodynamic limit is not well definite. 
Yet, we do not encounter special difficulties in characterizing this regime within our numerical studies at finite $N$.
In particular,
the ground state for $\theta=-\pi/2$ is the same for every value of $\alpha \geq 0$.
Furthermore, as discussed below, the QFI is superextensive in the FM phase (above and below $\alpha=1$) with the same scaling exponent.\\

{\it Antiferromagnetic phase --} For sufficiently strong AFM interaction, $\thetaAFM (\alpha) <\theta\leq\pi/2$, 
the system hosts an ordered AFM phase, 
where the staggered longitudinal magnetization  $\OP_z^{\mathrm{(st)}} = \bra{\psi_{\mathrm{gs}}}\hat{J}_z^{\rm (st)}\ket{\psi_{\mathrm{gs}}}$ acts as the  order parameter, with
$\hat{J}_z^{\rm (st)} =  \sum_{i=1}^N (-1)^i \hat{\sigma}_z^{(i)}$.
In particular, at $\theta\to\pi/2$, the ground state of a finite-size chain is the 
N\'eel state
$\ket{\psi_{\mathrm{gs}}} = [(\spinupz\spindownz)^{\otimes N/2} \allowbreak + \allowbreak (\spindownz\spinupz)^{\otimes N/2}]/\sqrt{2}$
for any $\alpha>0$. 
For $\alpha=0$, instead, each spin is coupled with all the others via the same strength, regardless of their mutual distance: 
the ground state at $\theta=\pi/2$ becomes the symmetric Dicke state (often also indicated as twin-Fock state) 
$\ket{\psi_{\mathrm{gs}}} = {\rm Sym}[\spinupz^{\otimes N/2} \spindownz^{\otimes N/2}]$,
given by the equally weighted superposition of all possible permutational symmetric combinations of $N/2$ spin-up and $N/2$ spin-down particles
(for an even number of spins). 
It should be noticed that for $\theta>0$, the energy of the ground state is extensive for all values of $\alpha \geq 0$, even for $\alpha=0$ and $\theta=\pi/2$.
This fact allows for a proper definition of the quantum phases even at $\alpha \leq 1$.\\

{\it Paramagnetic short-range and long-range phases --} A disordered paramagnetic (PM) phase is displayed by the system for weak spin-spin interaction, 
both in the FM and in the AFM regime, $\thetaFM (\alpha) <\theta<\thetaAFM (\alpha)$. 
The polarization provided by the transverse external magnetic field dominates over the spin-spin coupling and determines the structure of the ground state.
In particular, at $\theta=0$, the ground state is given by the coherent spin state 
$\ket{\psi_{\rm gs}} = (\spinupz
\allowbreak - \allowbreak 
\spindownz)^{\otimes N}/2^{N/2} 
\allowbreak = \allowbreak
\spindownx^{\otimes N}$ 
polarized along the $-x$ direction by the magnetic field.
In the following, we distinguish a paramagnetic SR phase from a LR one. 
This distinction is not based on an order parameter since the spin-flip $Z_2$ symmetry is preserved:
$\OP_z=0$ and  $\OP_z^{\rm (st)}=0$, in the full paramagnetic phase.
Instead, for $0< \theta < \thetaAFM(\alpha)$ and $\alpha \leq 1$, a logarithmic violation of the area law for the 
Von Neumann entropy has been found  in Ref.~\cite{KoffelPRL2012}, 
and shown not to originate from finite-size effects.
The analogy with critical gapless systems motivated the introduction of 
an effective central charge~\cite{KoffelPRL2012}, that has also been used as a tool for probing the phase diagram \cite{VodolaNJP2016}.
Finally, by means of a Jordan-Wigner transformation, the LR Ising chain can be mapped into a LR interacting fermionic chain \cite{VodolaNJP2016},
that, only in the PM regime at $\alpha \lesssim 1$, turns out to be characterized by the appearance of massive edge modes \cite{VodolaNJP2016}, 
similar to the ones found in the LR Kitaev chain \cite{VodolaPRL2014}.
All these peculiar features induce to conjecture the existence of a new PM phase at $\alpha \lesssim 1$ \cite{VodolaNJP2016}, 
bounded from above by a transition with nonvanishing mass gap at $\alpha \approx 1$.
This PM gapped phase, still preserving the $Z_2$ symmetry, 
will be quoted here and in the following as long-range paramagnetic (LRPM) phase, 
to distinguish it from the ordinary PM phase occurring at $\alpha\gtrsim1$.

In spite of the above theoretical clues, no valid observable for the experimental detection of the conjectured LRPM phase has been identified so far,  
mainly because BE is challenging to be observed in extended systems (see e.g. \cite{LewensteinBOOK, IslamNATURE2015}).
A similar open question holds for the nature of the AFM transitions at $\thetaAFM(\alpha)$: 
from the scaling of the Von Neumann entropy, the breakdown of conformal invariance induced by the LR interaction has been suggested \cite{VodolaNJP2016}. 
However, no detection criterion for observing the spontaneous breakdown of the conformal symmetry has been available so far to our knowledge. 
A promising method based on the inspection of the finite-size scaling of the ground-state energy density was suggested~\cite{VodolaPRL2014}, but its reliable use is currently forbidden by the limited size in experimental realizations of the LR Ising chain \cite{ZhangNATURE2017, BernienNATURE2017}.

\section{Multipartite-entanglement phase diagram}
\label{multipartite}

In order to characterize the phase diagram of the Hamiltonian (\ref{Hamiltonian}) 
beyond the analysis of order parameters and bipartite entanglement, 
we study here the QFI and its lower bound given by the spin-squeezing parameter.

The QFI of a generic state $\hat{\rho} = \sum_k p_k \vert k \rangle \langle k \vert$, relative to an arbitrary operator $\hat{O}$, is given by
(see the recent reviews \cite{Pezze2014, TothJPA2014, PezzeRMP} and references therein)
\be \label{QFImixed}
F_Q\big[\hat{\rho}, \hat{O}\big] = 2 \sum_{k,k'} \frac{(p_k - p_{k'})^2}{p_k + p_{k'}} \vert \langle k \vert  \hat{O} \vert k^{\prime} \rangle \vert^2,
\ee
in terms of eigenstates $\vert k \rangle$ and eigenvalues $p_k$ of the density matrix $\hat{\rho}$.
The QFI $F_Q[\hat{\rho}, \hat{O}]$ is related to the distinguishability between two nearby quantum states 
$\hat{\rho}$ and $\hat{\rho}(\phi) = e^{-i \hat{O} \phi} \hat{\rho} e^{i \hat{O} \phi}$ 
via the Uhlmann fidelity $\Tr[\sqrt{\hat{\rho}^{1/2}\hat{\rho}(\phi)\hat{\rho}^{1/2}}] = 1 - \frac{1}{8} F_Q[\hat{\rho}, \hat{O}] \phi^2 + \pazocal{O}(\phi^3)$:
the QFI thus quantifies the susceptibility of $\hat{\rho}$ to unitary parametric transformations.
For pure states $\ket{\psi}$, Eq.~(\ref{QFImixed}) reduces to the variance 
\be \label{QFIpure}
F_Q\big[\ket{\psi}, \hat{O}\big] = 4 \, \big( \bra{\psi}\hat{O}^2\ket{\psi} - \bra{\psi}\hat{O}\ket{\psi}^2 \big) \equiv 4 \,  (\Delta \hat{O})^2.
\ee
Notice that $F_Q[\hat{\rho}, \hat{O}]=0$ if and only if $[\hat{O}, \hat{\rho}] = 0$:
the QFI thus quantifies the ``spread'' of the state over the eigenstates of $\hat{O}$.

Importantly, the QFI is a witness of ME~\cite{HyllusPRA2012, TothPRA2012}.
Specifically, for collective operators $\hat{O} = \sum_i \hat{o}_i$ ($i$ labeling the lattice sites) the violation of the inequality
\be 
\label{QFIbound}
 f_Q[\hat{\rho}, \hat{O}] \equiv \frac{F_Q[\hat{\rho}, \hat{O}]}{N}  \leq k \, , 
\ee
signals $(k+1)$-partite entanglement ($1\leq k \leq N-1$) between spins~\cite{notaME}, where $f_Q$ is indicated as QFI density.
In particular, separable states $\hat{\rho}_{\rm sep}$ satisfy $f_Q\big[\hat{\rho}_{\rm sep}, \hat{O} \big] \leq 1$~\cite{PezzePRL2009}.
Moreover, states with $N-1 < f_Q[\hat{\rho}, \hat{O}] \leq N$ are genuinely $N$-partite entangled, 
$f_Q = N$ being the ultimate (Heisenberg) bound~\cite{PezzePRL2009,HyllusPRA2012, TothPRA2012}.

Here, we numerically study the QFI of the ground state $\ket{\psi_{\rm gs}}$ of the Hamiltonian~(\ref{Hamiltonian}) [see \ref{AppNumerical} for details on the numerical methods], 
with respect to the ordinary and staggered collective operators, $\hat{J}_{l} = \frac{1}{2} \sum_{i=1}^N \hat{\sigma}_{l}^{(i)}$ 
and $\hat{J}_{l}^{\rm (st)} = \frac{1}{2} \sum_{i=1}^N(-1)^i\hat{\sigma}_{l}^{(i)}$, respectively. 
We restrict ourselves to the two spatial directions $l=y,z$, since the direction $x$ provides $f_Q[\ket{\psi_{\rm gs}}, \hat{J}_x]\leq 1$ and 
$f_Q[\ket{\psi_{\rm gs}}, \hat{J}_x^{\rm (st)}] \leq 1$
for all values of the parameters $\theta$ and $\alpha$.
A central step in this calculation is the relation
between the QFI relative to the collective operators 
and the connected correlation functions
$C_{ll}^{(i,j)}=\bra{\psi_{\mathrm{gs}}}\hat{\sigma}^{(i)}_l \hat{\sigma}^{(j)}_l \ket{\psi_{\mathrm{gs}}} - \bra{\psi_{\mathrm{gs}}}\hat{\sigma}^{(i)}_l \ket{\psi_{\mathrm{gs}}} \bra{\psi_{\mathrm{gs}}}\hat{\sigma}^{(j)}_l \ket{\psi_{\mathrm{gs}}}$ 
\cite{HaukeNATPHYS2016, PezzePRL2017}:
\be \label{QFIcorr}
f_Q\big[\ket{\psi_{\mathrm{gs}}},\hat{J}_l\big] = \frac{1}{N} \sum_{i,j=1}^N C_{ll}^{(i,j)} 
\ee
and
\be
f_Q\big[\ket{\psi_{\mathrm{gs}}},\hat{J}_l^{\rm (st)}\big] = \frac{1}{N} \sum_{i,j=1}^N (-1)^{i-j} C_{ll}^{(i,j)}.
\ee
It should be noticed that different operators $\hat{O}$ yield different values of the QFI. 
The calculation of the QFI for different operators provides a ``tomographic survey of ME'' for the given quantum states 
[in particular of the ground state of the Hamiltonian (\ref{Hamiltonian})] that is able, as illustrated below, 
to fully characterize the phase diagram.\\

We also analyze the Wineland spin-squeezing (WSS) parameter~\cite{WinelandPRA1994,MaPHYSREP2011,PezzeRMP}
 \be \label{WSS} 
\xi^2_{\rm R} = \frac{N(\Delta\hat{J}_{\mathbf{n}_\perp})^2}{\langle\hat{J}_{\mathbf{n}_\parallel}\rangle^2},
\ee
defined in terms of first and second momenta of the collective spin operators $\hat{J}_{l}$. 
In Eq.~(\ref{WSS}),  $\mathbf{n}_\parallel$ and $\mathbf{n}_\perp$ are orthogonal directions chosen in order to minimize $\xi^2_{\rm R}$.
A state is said to be spin squeezed along the direction $\mathbf{n}_\perp$ if $\xi^2_{\rm R}<1$.
This inequality is also a criterion for entanglement~\cite{SorensenNAT2001} and has been extended to witness ME~\cite{SorensenPRL2001}. 
The inverse of the spin-squeezing parameter (\ref{WSS}) is a lower bound of the QFI~\cite{TothJPA2014, PezzeRMP, Pezze2014}: for 
any state $\hat{\rho}$ we have $N/\xi_{\mathrm{R}}^2 \leq F_Q[\hat{\rho}, \hat{J}_{\mathbf{n}'_\perp}]$, 
where $\mathbf{n}'_\perp$ is a direction orthogonal to both $\mathbf{n}_\perp$ and $\mathbf{n}_\parallel$. 
Notice that for pure states the inequality $N/\xi_{\mathrm{R}}^2 \leq F_Q[\ket{\psi}, \hat{J}_{\mathbf{n}'_\perp}] = 4(\Delta \hat{J}_{\mathbf{n}'_\perp})^2$ 
follows from the Heisenberg uncertainty relation.
The spin squeezing is also related to the correlation function of collective spin operators and, for finite $\langle\hat{J}_{\mathbf{n}_\parallel}\rangle$, 
has the scaling properties of $(\Delta\hat{J}_{\mathbf{n}_\perp})^2$. 

The investigation of ME in the ground state of the Ising chains, as witnessed by the QFI and the WSS, has been limited  so far to the two extreme cases of nearest-neighbor 
$\alpha = \infty$~\cite{LiuJPA2013,HaukeNATPHYS2016,FrerotPRL2018} and infinite-range $\alpha=0$ coupling \cite{MaPRA2009,FrerotPRL2018}.
Several works have analyzed the QFI and the WSS in the ground state of the bosonic Josephson junction, which formally corresponds to
the fully connected Ising model restricted to the Hilbert subspace of states that are symmetric under particle exchange~\cite{PezzePRA2005, HaukeNATPHYS2016, PezzeRMP}:
see Refs.~\cite{EsteveNATURE2008, BerradaNATCOMM2013, TrenkwalderNATPHYS2016, PezzeRMP} for experimental investigations in Bose-Einstein condensates. 
Notice that the ground state of the Hamiltonian~(\ref{Hamiltonian}) for $\alpha=0$ is indeed given by symmetric states.  \\

\begin{figure*}[!t]
\centering
\includegraphics[width=1.1\textwidth]{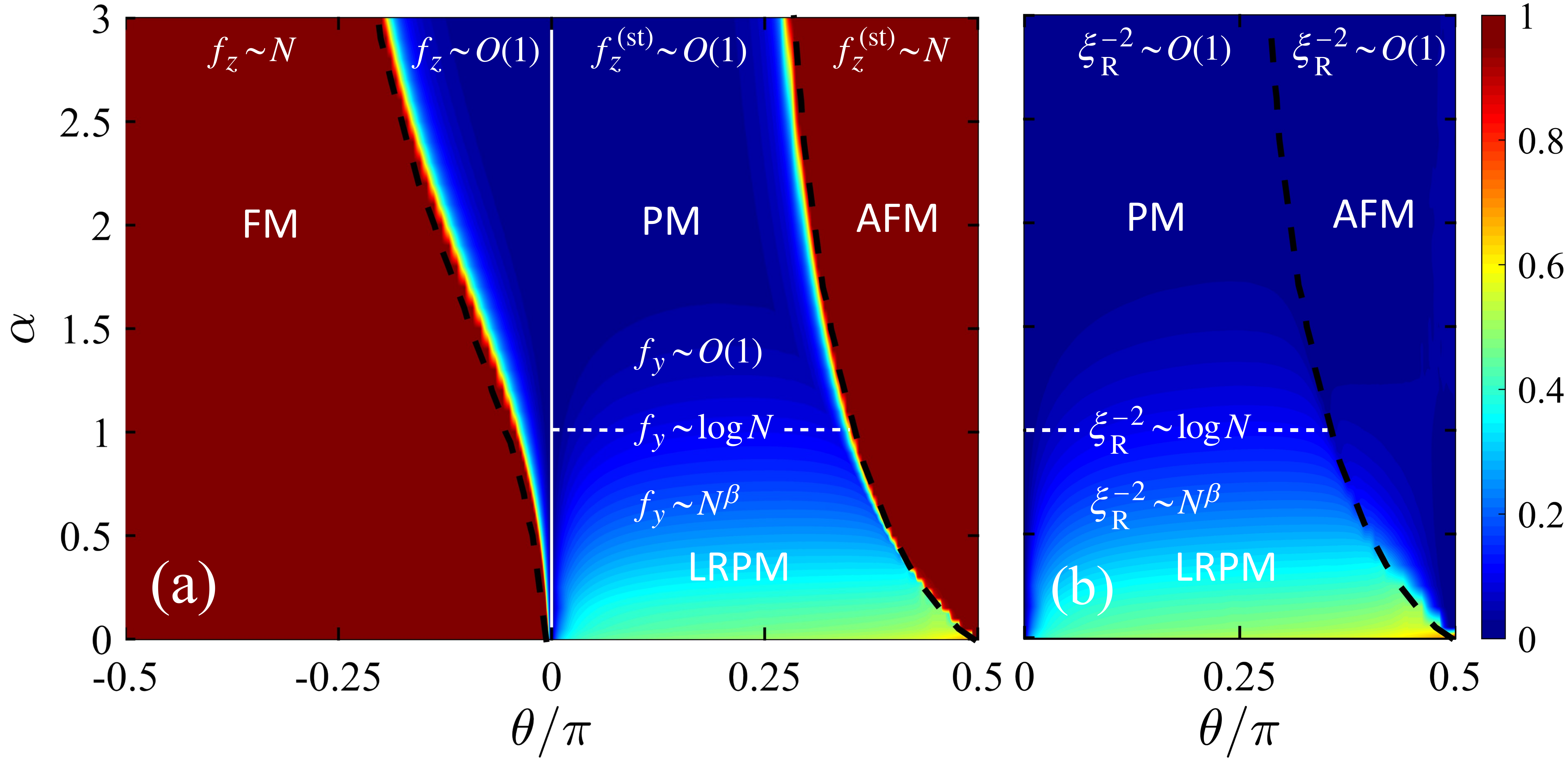} 
\caption{
Scaling power of the QFI density, $d \log f_Q / d \log N $ (left panel, color scale), 
and of the inverse spin squeezing parameter, $d \log \xi^{-2}_{\rm R} / d \log N$ (right panel, color scale),
for the ground states of the Ising chain (\ref{Hamiltonian}) on the $\theta$-$\alpha$ plane.
The black dashed lines mark the minimum of the mass gap. 
The vertical white line corresponds to $\theta=0$, where $f_Q = 1$.
The asymptotic scaling of the QFI with $N$ is highlighted in the different regions, using the 
compact notation $f_{z,y} \equiv f_Q[\ket{\psi_{\mathrm{gs}}},\hat{J}_{z,y}]$ and
$f_z^{\rm (st)} \equiv f_Q[\ket{\psi_{\mathrm{gs}}},\hat{J}_z^{\rm(st)}]$, while the 
spin squeezing parameter is calculates as $\xi^{2}_{\rm R} = N (\Delta \hat{J}_z)^2/\mean{\hat{J}_x}^2$.
In both panels $N=50$ spins.}
\label{fig2}
\end{figure*}

In the following we provide a study of the model (\ref{Hamiltonian}) in the full range $0 \leq \alpha \leq \infty$. 
We find that the QFI witnesses ME, $f_Q  >1$, for any $\alpha\geq 0$ and $\theta \neq 0$,
when calculated with respect to the optimal operators reported in Fig.~\ref{fig2}(a).
Instead, on the line $\theta=0$, the ground state is separable (for any $\alpha$),
and the QFI does not overcome the bound $f_Q=1$.
In the PM phase for $\theta>0$, ME is also witnessed by the spin-squeezing parameter, as shown in Fig.~\ref{fig2}(b). 
We point out that, while the figure is obtained at $N=50$, we have checked the qualitative stability of the phase diagram as $N$ increases up to $N \approx 200$: 
the change of behavior around $\alpha=1$ becomes sharper.

\subsection{Ferromagnetic regime}
\begin{figure*}[!t]
\centering
\includegraphics[width=0.9\textwidth]{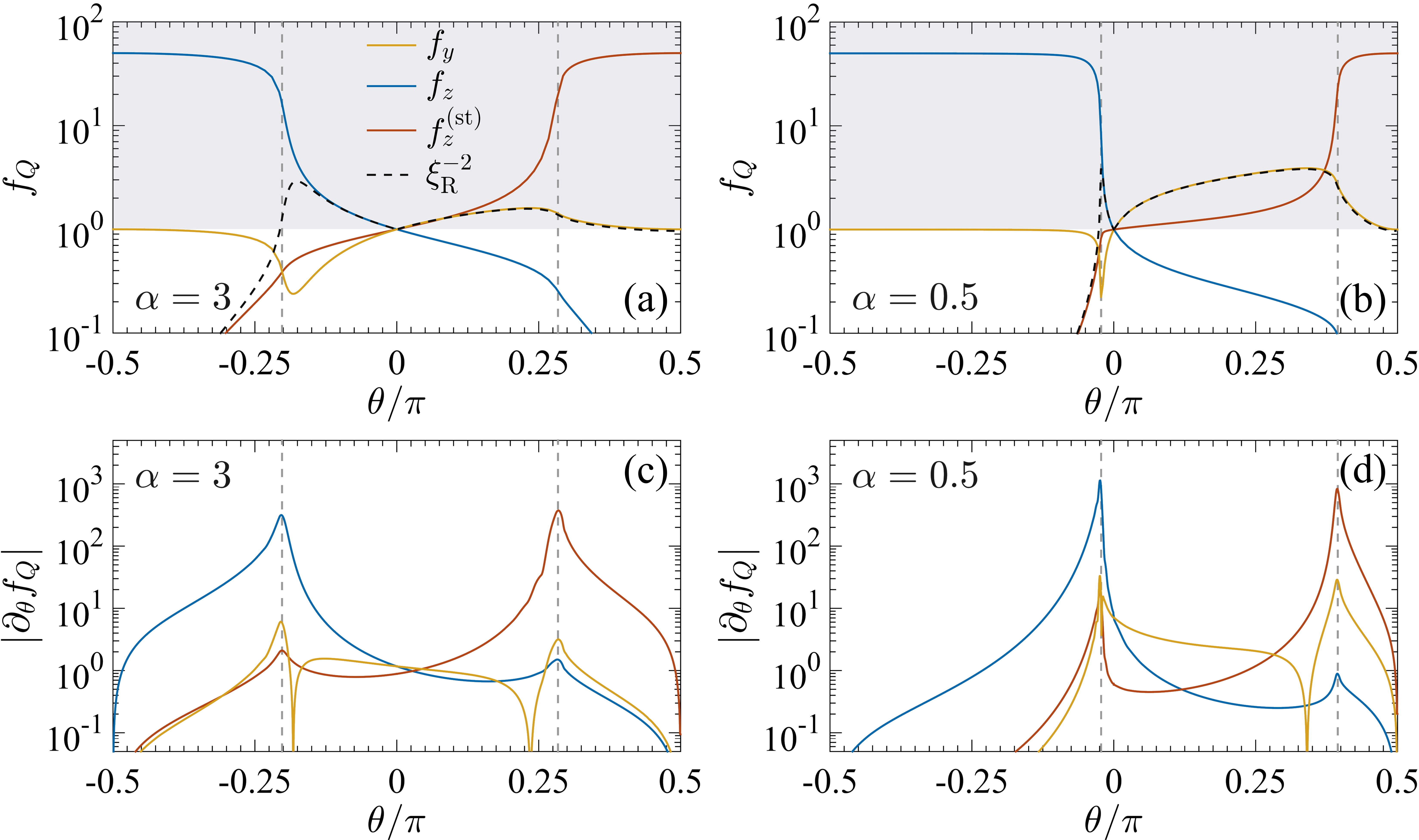} 
\caption{Panels (a) and (b): the solid lines show the QFI density $f_{z,y} \equiv f_Q[\ket{\psi_{\mathrm{gs}}},\hat{J}_{z,y}]$ and
$f_z^{\rm (st)} \equiv f_Q[\ket{\psi_{\mathrm{gs}}},\hat{J}_z^{\rm(st)}]$ as a function of $\theta$.
The black dashed line is the inverse spin-squeezing parameter calculates as $\xi_{\rm R}^2 = N (\Delta \hat{J}_z)^2/\mean{\hat{J}_x}^2$ for $\theta\geq 0$
and $\xi_{\rm R}^2 = N (\Delta \hat{J}_y)^2/\mean{\hat{J}_x}^2$ for $\theta < 0$.
The vertical grey dashed lines indicate the position of the critical points, extrapolated in the limit $N\to\infty$. 
Values of the QFI density in the grey region (corresponding $f_Q>1$) are only possible for entangled states.
In panel (c) and (d) we plot the derivative of $f_{y,z}$ and $f_z^{\rm(st)}$ with respect to $\theta$.
Here $N=50$, panels (a) and (c) refer to $\alpha=3$, while (b) and (d) to $\alpha=0.5$.}
\label{fig3}
\end{figure*}

For $\theta<0$ the QFI is maximized when calculated with respect to the operator $\hat{J}_z$, which is the order parameter of the PM-to-FM quantum phase transitions, see Fig.~\ref{fig3}.
In the FM phase, $\theta<\thetaFM(\alpha)$, we find the power-law scaling $f_Q[\ket{\psi_{\rm gs}},\hat{J}_z] \sim N$ for any $\alpha$, with a prefactor that depends on $\theta$.
In particular, at $\theta\to-\pi/2$, where the ground state is given by the GHZ state,
the Heisenberg limit $f_Q[\ket{\psi_{\mathrm{gs}}},\hat{J}_z] = N$ is recovered. 
It should be noticed that $f_Q[\ket{\psi_{\rm gs}},\hat{J}_z] \sim N$ in the FM phase both above and below $\alpha=1$ 
despite the superextensive energy scaling in the LR regime.
Conversely, the QFI is only extensive in the PM phase, $f_Q[\ket{\psi_{\mathrm{gs}}},\hat{J}_z] \sim \pazocal{O}(1)$.
Still, the QFI witnesses ME: we find $f_Q[\ket{\psi_{\mathrm{gs}}},\hat{J}_z] > 1$ in the full PM phase.
The PM-to-FM quantum phase transition at $\thetaFM(\alpha)$ marks a change of scaling of the QFI with $N$. 
The derivative of the QFI with respect to $\theta$, $d f_Q[\ket{\psi_{\mathrm{gs}}},\hat{J}_z]/ d \theta$ is thus 
characterized by a pronounced maximum at $\theta=\theta^-_{c}(\alpha)$, see Fig.~\ref{fig3}, 
that diverges in the thermodynamic limit.

\subsection{Antiferromagnetic regime}
\begin{figure*}[!t]
\centering
\includegraphics[width=0.8\textwidth]{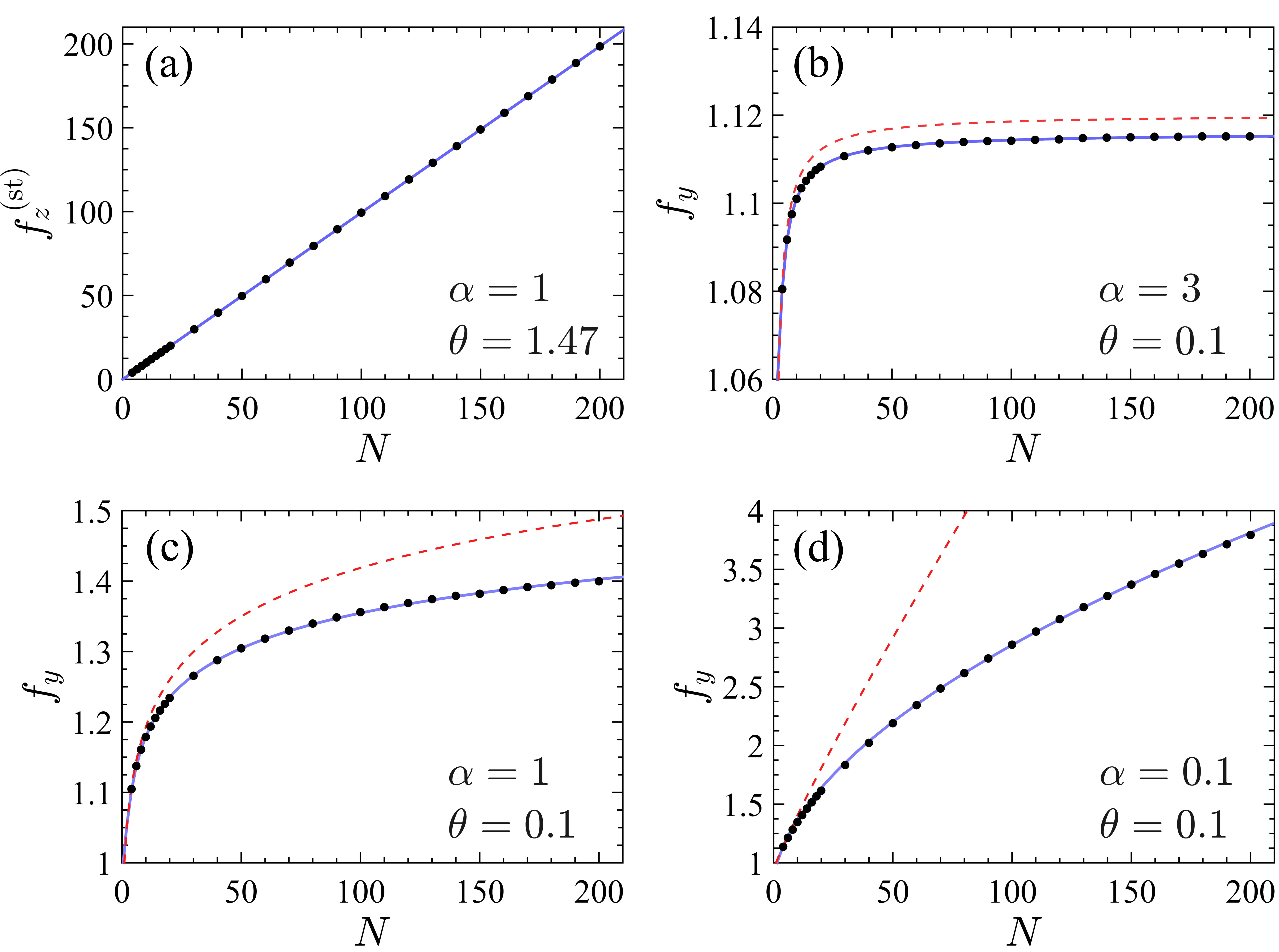} 
\caption{Finite-size scaling of the QFI (dots) with $N$. Different panels are:
(a) $\alpha=1$ and $\theta=1.47$ (AFM phase); 
(b) $\alpha=3$ and $\theta=0.1$ (PM phase);
(c) $\alpha=1$ and $\theta=0.1$ (transition between PM and LRPM phases);
(d) $\alpha=0.1$ and $\theta=0.1$ (LRPM phase). 
The blue solid lines are fits: 
(a) $f_z^{\rm (st)} = 0.99 N^{1.00}$
(b) $f_y = 1.12 - 0.13/N^{0.94}$, 
(c) $f_y = 1 + 0.08\log N$, 
(d) $f_y = 1 + 0.17 N^{0.54}$~\cite{notaLL}.
In all panels, the red dashed lines are analytical predictions obtained with a perturbative approach and valid for sufficiently small $N$, 
see \ref{AppPerturb}.}
\label{fig4}
\end{figure*}

The AFM regime is richer than the FM one. 
In the AFM phase, for $\theta > \thetaAFM(\alpha)$ and $\alpha>0$, 
the QFI is maximized when calculated with respect to $\hat{O}=\hat{J}_z^{\rm (st)}$, which is the order parameter of the PM-to-AFM quantum phase transition. 
Similarly as above, this QPT is associated to a divergence of the derivative of the QFI with respect to $\theta$, 
$d f_Q[\ket{\psi_{\mathrm{gs}}},\hat{J}_z^{\rm (st)}] / d \theta$, see Fig.~\ref{fig3}.
In the AFM phase, the QFI has a superextensive scaling: we find $f_Q[\ket{\psi_{\mathrm{gs}}},\hat{J}_z^{\rm (st)}] = c(\alpha,\theta) \, N$  with $c(\alpha,\theta)\leq1$.
In particular, for $\alpha=\infty$, the analytical calculation of the correlation functions \cite{Lieb1961} provides $c(\infty, \theta)=(1-\cot^2\theta)^{1/4}$.
In the limit $\theta\to\pi/2$, where  the ground state is the N\'eel state,
the Heisenberg limit $c(\alpha,\theta)=1$ is strictly saturated for all values of $\alpha>0$, see Fig.~\ref{fig4}(a) 
for a plot of $f_Q[\ket{\psi_{\mathrm{gs}}},\hat{J}_z^{\rm (st)}]$ as a function of $N$ in the AFM phase.
At $\alpha=0$ the ground state is instead given by the symmetric Dicke state and we have $c(\alpha,\theta)=1/2+1/N$.

In the paramagnetic phase, $0 < \theta < \thetaAFM(\alpha)$, the QFI has two clearly distinguished behaviors, see Figs.~\ref{fig2} and \ref{fig3}. 
For SR coupling, $\alpha>1$, 
we find an extensive QFI, $f_Q[\ket{\psi_{\mathrm{gs}}},\hat{J}_z^{\rm (st)}] \sim \pazocal{O}(1)$ and $f_Q[\ket{\psi_{\mathrm{gs}}},\hat{J}_y] \sim \pazocal{O}(1)$, see Fig.~\ref{fig4}(b): 
the quadratic term in the Hamiltonian (\ref{Hamiltonian}) is responsible for ME ($f_Q>1$), but the entanglement depth does not scale with the system size. 
In particular, for $\alpha=\infty$, 
the QFI is maximized when calculated with respect to $\hat{O}=\hat{J}_z^{\rm (st)}$ for all values of $0 < \theta < \thetaAFM(\alpha)$.
There, the correlation function $C_{zz}^{(i,j)} \sim (-1)^{i-j} e^{-|i-j|/\xi}$ induces 
$f_Q[\ket{\psi_{\mathrm{gs}}}, \hat{J}_z^{\rm (st)}] \sim 2 (1-e^{-1/\xi})^{-1}$, in virtue of Eq. (\ref{QFIcorr}), where $\xi$ is the (finite) correlation length.
On the contrary, in the LRPM phase at $\alpha \leq 1$,  
the QFI is maximized by $\hat{O}=\hat{J}_y$, that is not the order parameter of the PM-to-AFM quantum phase transition.
Here, the QFI has a superextensive scaling.
For $\alpha=1$, we find the logarithmic behavior $f_Q[\ket{\psi_{\mathrm{gs}}},\hat{J}_y] \sim \log{N}$ analytically suggested by 
a perturbative calculation, see \ref{AppPerturb}, and
tested by numerical calculations up to $N=200$, see Fig. \ref{fig4}(c).
For $\alpha<1$ we find a power-law behavior $f_Q[\ket{\psi_{\mathrm{gs}}},\hat{J}_y] \sim N^{\pow(\alpha)}$, 
where $0< \pow(\alpha) \leq 0.5$, see Fig. \ref{fig4}(d).
In particular, a variational ansatz at $\alpha=0$ predicts $f_Q[\ket{\psi_{\mathrm{gs}}},\hat{J}_y] = \sqrt{N \tan \theta}$, see \ref{AppVariational}, 
in very good agreement with the numerical calculations for large $N$.

The super-extensiveness of the QFI directly stems from the power-law tail in the algebraic decay 
of the correlation functions $C_{yy}^{(i,j)}$~\cite{KoffelPRL2012,VodolaNJP2016}.
Interestingly, the behavior of the QFI in the PM phase is fully captured by the spin-squeezing parameter: we find 
$\xi^{-2}_{\rm R} = \mean{ \hat{J}_x }^2 / (N (\Delta \hat{J}_z)^2) \approx f_Q[\ket{\psi_{\rm gs}}\hat{J}_y]$, as shown in Figs.~\ref{fig2} and~\ref{fig3}(b).

\begin{figure*}[!t]
\centering
\includegraphics[width=1\textwidth]{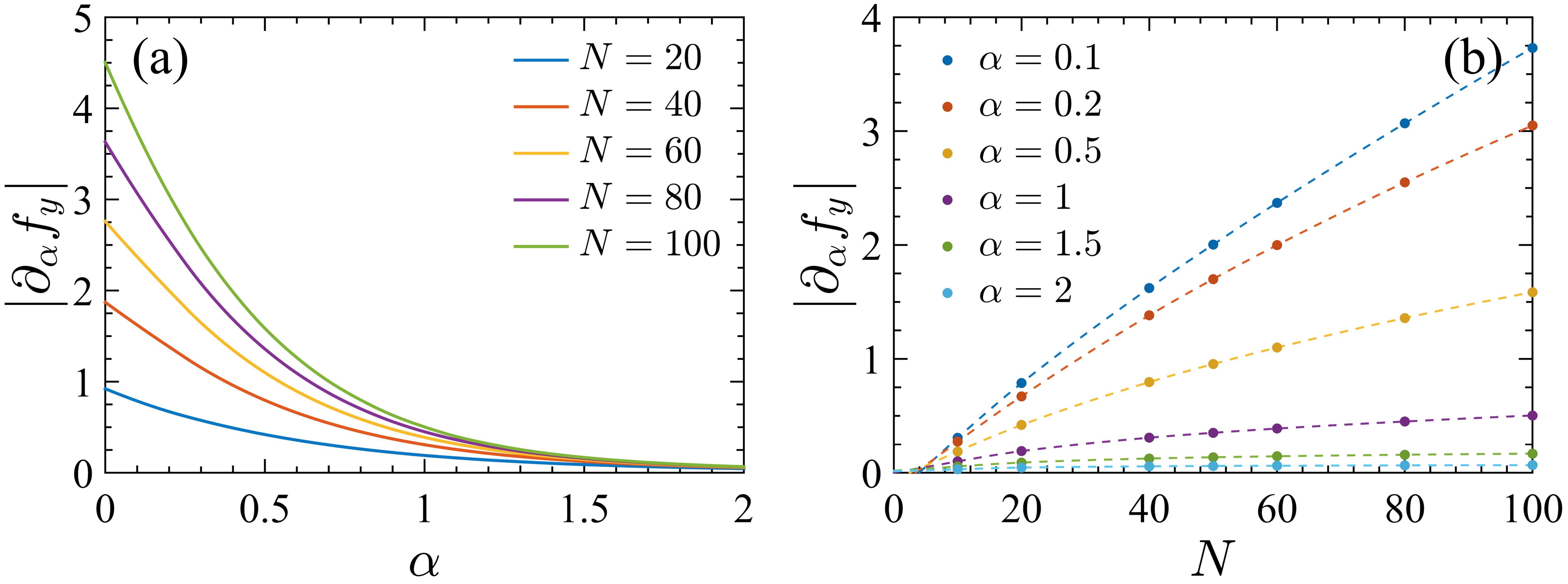} 
\caption{Derivative of $f_Q[\ket{\psi_{\rm gs}}, \hat{J}_y]$ with respect to $\alpha$ as a function of $\alpha$ (a) and $N$ (b) for $\theta = 0.1$.}
\label{fig5}
\end{figure*}

\subsection{Crossing the massive line $\alpha=1$ in the antiferromagnetic regime}

As discussed above, and shown in Fig.~\ref{fig2}, 
when crossing the massive line $\alpha=1$, the scaling of the QFI  
with $N$ changes from extensive (for $\alpha>1$) to superextensive (for $\alpha\leq 1$).
This result can be taken as a strong indication for a gapped QPT occurring at $\alpha=1$ from a SR phase to a LR phase. 
This is a consequence of the change of behavior of the correlation function that is captured by the QFI. 
More explicitly, assuming $f_y \equiv f_Q[\ket{\psi_{\mathrm{gs}}},\hat{J}_y] \approx a(\alpha)\times N^{\beta(\alpha)}$, as obtained from our numerics, 
we find 
\be
\frac{d f_y}{d \alpha} = N^{\beta(\alpha)} \Big( \frac{d a(\alpha)}{d \alpha} + a(\alpha) \frac{d \beta(\alpha)}{d \alpha} \log N \Big).
\ee
If $\beta(\alpha) =0$, as obtained for $\alpha>1$, we find that $\frac{d f_y}{d \alpha} = \frac{d a(\alpha)}{d \alpha}$ does not scale with $N$.
Conversely, if $\beta(\alpha) \neq 0$, as obtained for $\alpha<1$, we find $\frac{d f_y}{d \alpha} \approx N^{\beta(\alpha)}$.
In Fig.~\ref{fig5} we plot $\frac{d f_y}{d \alpha}$, obtained numerically (without any assumption on the the functional form of $f_y$), 
as a function of $\alpha$ [panel (a)] and as a function of $N$ [panel (b)]. 
Both panels suggest (despite the system size limited to $N=100$) a sharp change of behavior around $\alpha=1$:
while $\frac{d f_y}{d \alpha}$ increases with $N$ for $\alpha\leq 1$, it remains approximately constant for $\alpha>1$.

To gain more insight into the behavior of $\frac{d f_y}{d \alpha}$ at large $N$
we consider the results of the perturbative calculation for $\theta \to 0^+$, see \ref{AppPerturb}: we have
\be \label{DerPerturbationFunction}
\frac{d f_Q[\ket{\psi_{\rm gs}}, \hat{J}_y]}{d \alpha} = \sqrt{8 \, \frac{N-1}{N}}\,\theta \, \frac{d \mathcal{G}_N(\alpha)}{d\alpha}, 
\ee
where, to leading order in $N$,
\be \label{scalingQFIpert}
\frac{d \mathcal{G}_N(\alpha)}{d\alpha} \approx 
\left\{
\begin{array}{ll} 
\frac{d \zeta(\alpha)}{d \alpha} & {\rm for} \ \alpha>1 \\ 
\frac{1}{(\alpha-1)(\alpha-2)}\,N^{1-\alpha}\log{N} & {\rm for} \ 0<\alpha<1
\end{array}
\right.
\ee
$\zeta(\alpha)$ being the Riemann zeta function.
This analysis supports the numerical findings: $\frac{d f_y}{d \alpha}$ increases with $N$ for $\alpha < 1$, while it does not scale with $N$ for $\alpha > 1$.
A similar behavior as in Eq.~(\ref{DerPerturbationFunction}) is revealed by the fidelity susceptibility, 
again obtained from a perturbative calculation, see \ref{AppPerturb}.
It should be noticed, however, that the condition of validity of perturbation theory, $\,\theta\,\mathcal{G}_N(\alpha)\ll1$, sets an upper limit 
for the validity of Eq.~(\ref{DerPerturbationFunction}): 
for fixed $\theta\ll1$, the finite-size scaling for $\alpha<1$ is only guaranteed when $N\ll\theta^{-1/(1-\alpha)}$ . 
Namely, from Eq.~(\ref{DerPerturbationFunction}) we cannot claim a superextensive scaling in the thermodynamic limit.

Summing up, our numerical and analytical results allow to locate the boundary between the SR and LR regimes at $\alpha=1$, also improving the precision of previous studies~\cite{VodolaNJP2016}.

\begin{figure*}[!t]
\centering
\includegraphics[width=1.1\textwidth]{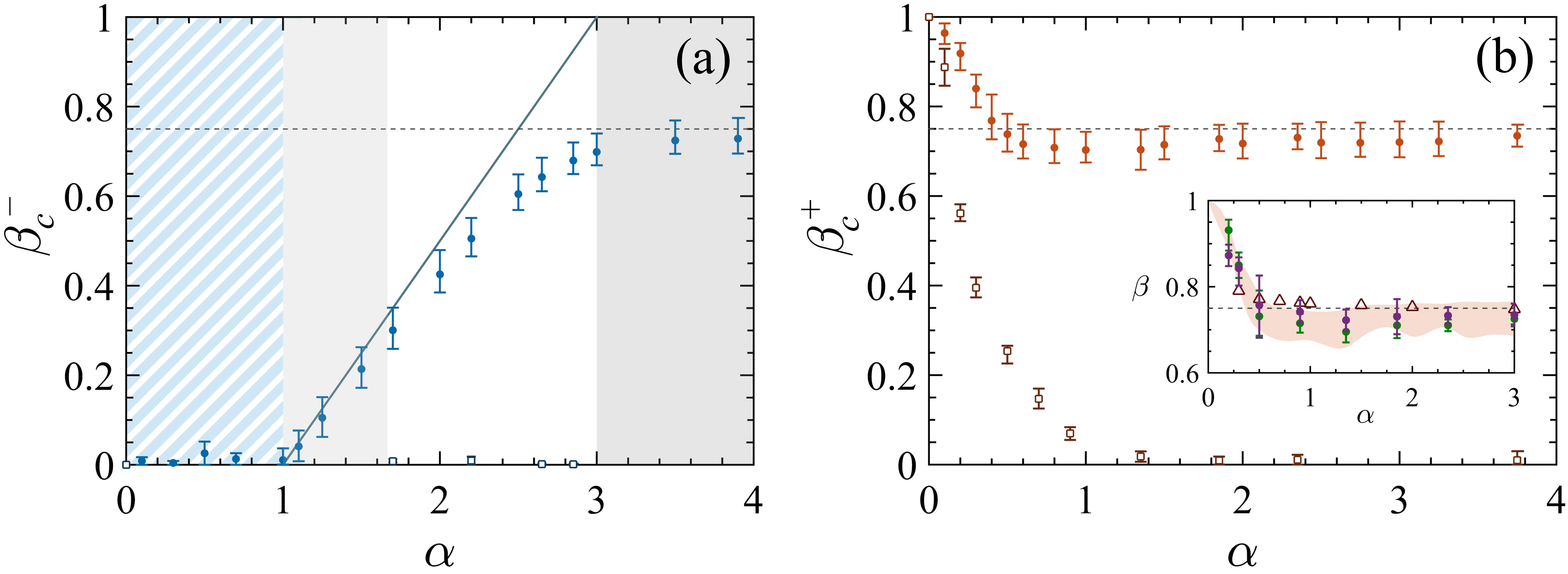} 
\caption{Scaling power for the QFI density,  $\pow_c^{\pm}(\alpha) = d\log f_Q / d \log N$, along the critical massless lines. 
(a) Scaling $\pow_c^-(\alpha)$ along $\thetaFM(\alpha)$ (blue dots).
In the hatched region the thermodynamic limit is not definite:
here $\pow_c^-(\alpha) \approx 0$. 
The solid blue line is the mean-field prediction $\pow_c^-(\alpha) = (\alpha-1)/2$, which is expected to be accurate for $1< \alpha< 5/3$ (light grey region).
For $\alpha \gtrsim 3$ (dark grey region) the scaling for the nearest-neighbor Ising model $\pow_c^-=3/4$ (horizontal dashed line) is expected to hold.   
Open squares are the scaling of the spin-squeezing parameter $d \log \xi^{-2}_{\rm R} / d \log N$. 
(b) Scaling $\pow_c^+(\alpha)$ along $\thetaAFM(\alpha)$ (red dots) 
as a function of $\alpha$. 
The dashed line is $\beta=3/4$.
Open squares are the scaling of the spin-squeezing parameter $d \log \xi^{-2}_{\rm R} / d \log N$.
The inset shows the scaling power $\pow(\alpha)$ obtained directly from power-law fits on the correlation functions 
$\langle\hat{\sigma}_{z}^{(N/2)}\hat{\sigma}_{z}^{(N/2+N/5)} \rangle \sim N^{\beta(\alpha)-1}$ as a function of $N$ (green dots)
and $\langle\hat{\sigma}_{z}^{(N/2)}\hat{\sigma}_{z}^{(N/2+r)} \rangle \sim r^{\beta(\alpha)-1}$ as a function of $r$ at $N=120$ (purple circles).
These results are consistent, within error bars, with the values of $\pow_c^+(\alpha)$ obtained from the analysis of the QFI 
(orange region, corresponding to the values shown in the main figure).
For comparison, we report the results of Ref.~\cite{KoffelPRL2012} (red triangles).
In both the panels, the error bars are due to the uncertainty on the location of the critical points.
}
\label{fig6}
\end{figure*}

\section{Quantum Fisher information along the massless critical lines}

The QFI  is also useful to probe directly conformal invariance along the critical 
lines $\thetac^{\pm}(\alpha)$, see Fig.~\ref{fig5}. 
Indeed, the QFI density $f_Q[\ket{\psi_{\mathrm{gs}}},\hat{O}]$ 
($\hat{O}$ being here the order parameter of the transition) 
scales with the systems size at criticality as  $f_Q[\ket{\psi_{\mathrm{gs}}},\hat{O}] \sim N^{d- 2 \, \Delta_{\hat{O}}}$ ($d = 1$ in our  case),
where $\Delta_{\hat{O}}$ is the scaling dimension of $\hat{O}$ \cite{HaukeNATPHYS2016}. 
At criticality and for one-dimensional quantum systems, conformal invariance fully constrains the set of possible $\Delta_{\hat{O}}$ (see e.g. \cite{yellowbook,mussardo}).

For the AFM transition, we probe the scaling of $f_Q[\ket{\psi_{\mathrm{gs}}},\hat{J}_z^{\rm (st)}]$ with the systems size $N$ along $\thetaAFM (\alpha)$, 
while for the FM transition we probe the scaling of $f_Q[\ket{\psi_{\mathrm{gs}}},\hat{J}_z]$ along $\thetaFM(\alpha)$.
In both cases, such a scaling is known to be constrained by conformal invariance
to $N^{3/4}$, corresponding to $\Delta_{\hat{O}}= 1/8$~\cite{Sachdev}.

Figure~\ref{fig5} shows 
\be
\pow_c^{\pm}(\alpha) = \frac{d \log f_Q}{d \log N} \bigg\vert_{\theta = \thetac^{\pm}(\alpha)}
\ee 
as a function of $\alpha$ along $\thetaFM(\alpha)$ [panel (a)] and $\thetaAFM(\alpha)$ [panel (b)], as determined numerically
from a finite-size analysis of our data for $N=10\dots120$. 
The error bars are mainly due to the numerical indeterminacy in finding the critical point $\theta^{\pm}_c(\alpha)$, identified here as the minimum of the mass gap.\\

{\it Scaling along the FM critical line.} Along $\thetaFM(\alpha)$, the conformal scaling holds for $\alpha \gtrsim 3$, see Fig.~\ref{fig5}(a).
This is consistent with the results of Ref.~\cite{DuttaPRB2001} where it has been shown that for $\alpha \geq 3$
the phase transition is in the universality class of the short-range Ising transition, see also Refs.~\cite{KnapPRL2013, DefenuPRB2017, ZhuARXIV2018}.
For $\alpha \lesssim 3$, we find that $\pow(\alpha)$ decreases down to $\pow=0$ at $\alpha=1$.

The scaling of the QFI density for $\alpha \gtrsim 1$ can be obtained from a Landau-Ginzburg approach. 
We introduce the effective action 
\be \label{lglagr}
S =  \int \mathrm{d} x \, \mathrm{d} t \, \Bigg[\phi^{\dagger}(x, t) \Big(-\partial_{t}^2 + \partial_x^{(\alpha -1)} \Big) \phi(x,t) +g \, |\phi(x,t)|^4  + o\big(|\phi(x,t)|^4\big) \Bigg],
\ee
where  $\phi (x, t)$ represents,  at the low-energy effective level,  the order parameter  $\OP_z = \bra{\psi_{\mathrm{gs}}}\hat{J}_z\ket{\psi_{\mathrm{gs}}}$, 
$g$ is a coupling constant, and $\partial_x^{\gamma}$ denotes the fractional derivative. 
This action can be justified by a renormalization group procedure, suitably modified for LR models \cite{LeporiAP2016,LeporiJSM2016}, and it is known to be dominant, in the range $1 \leq \alpha \lesssim 3$ \cite{Sachdev,KnapPRL2013}, with respect to the conformal Landau-Ginzburg action \cite{zamLG,mussardo}
\be
S =  \int \mathrm{d} x \, \mathrm{d} t \, \Bigg[\phi^{\dagger}(x, t) \Big(- \partial_{t}^2 + \partial_x^2 \Big) \phi(x,t) +g \, |\phi(x,t)|^4  +o\big(|\phi(x,t)|^4\big) \Bigg].
\label{lglagr2}
\ee
The action (\ref{lglagr}) predicts  the breakdown of the conformal invariance [owned instead by (\ref{lglagr2})]~\cite{LeporiJSM2016}:
for instance, beyond the Lorentz (Euclidean) rotational invariance, the invariance under dilatations $(t, x) \to \lambda \, (t, x)$ is lost, substituted by an ``asymmetric'' version counterpart
$(t, x) \to (\lambda \, t, \lambda^{\frac{2}{\alpha -1}} \,  x)$. 
This fact is also associated to an anomalous
dynamical exponent $z = \frac{\alpha-1}{2}$ \cite{KnapPRL2013}. More importantly,  (\ref{lglagr}) implies
the behavior for the time-independent correlations \cite{MaghrebiPRB2013}
\be
\bra{\psi_{\mathrm{gs}}} \phi (0, 0) \phi (x, 0) \ket{\psi_{\mathrm{gs}}} \sim \frac{1}{x^{1-\frac{\alpha -1}{2}}} \, . 
\label{anscal}
\ee
Exploiting the relation~(\ref{QFIcorr}) between the QFI and the two-points correlation functions, 
we have
\be \label{Fishmf}
f[\ket{\psi_{\mathrm{gs}}},\hat{J}_z] \sim N^{\frac{\alpha -1}{2}}
\ee 
giving $\beta_c^-(\alpha) = (\alpha -1)/2$.
This result agrees well with our numerical calculations, see Fig. (\ref{fig5}). 
Equation~(\ref{Fishmf}) is also recovered taking into account the relation $\Delta_{\hat{O}}=-(1-\eta-z)/2$, and using the mean-field critical exponents 
$\eta^{\rm mf} = 3-\alpha$ and $z^{\rm mf} = (\alpha-1)/2$ calculated in Ref.~\cite{KnapPRL2013}, giving $\beta_c^{-,{\rm mf}}(\alpha) = 1-2\Delta_{\hat{O}}^{\rm mf}=2-\eta^{\rm mf}-z^{\rm mf} = (\alpha-1)/2$.
This prediction is expected to be accurate for $\alpha < 5/3$~\cite{KnapPRL2013}.
For larger values of $\alpha$, the deviation of the scaling of the QFI from Eq.~(\ref{Fishmf}) is 
probably a clue that a more careful renormalization group treatment is required approaching $\alpha = 3$, such to properly account
 the interplay between (\ref{lglagr}) and (\ref{lglagr2}).
In Fig.~\ref{fig5} we also show the scaling of the spin-squeezing parameter $\xi_{\rm R}^2 = N (\Delta \hat{J}_y)^2/\mean{\hat{J}_x}^2$ at $\thetaFM(\alpha)$. 
We find $d \log \xi_{\rm R}^{-2}/d\log N \approx 0$ for all values of $\alpha$: differently from the QFI, $\xi_{\rm R}^{-2}$ does not scale at the transition point. \\

{\it Scaling along the AFM critical line.} Along $\thetaAFM(\alpha)$ we find $\pow_c^+(\alpha) \approx 3/4$ for $\alpha \gtrsim 0.5$, see Fig.~\ref{fig5}(b).
For  $\alpha \lesssim 0.5$, $\pow_c^+$ increases smoothly up to $\pow_c^+=1$ at $\alpha = 0$.
Notice that the scaling 
$f_Q[\ket{\psi_{\mathrm{gs}}},\hat{J}_z^{\rm (st)}] \sim N$ at $\thetaAFM(0) = \pi/2$ is analytically known and recovered by our numerics.
In Fig.~\ref{fig5}(b) we also report the scaling of the spin-squeezing parameter $\xi_{\rm R}^2 = N (\Delta \hat{J}_z)^2/\mean{\hat{J}_x}^2$. 
We find $d \log \xi_{\rm R}^{-2}/d\log N \approx 0$ for $\alpha \gtrsim 1$, while it increases for $\alpha \lesssim 1$.

In Ref.~\cite{KoffelPRL2012}  the scaling of $\langle\hat{\sigma}_{z}^{(N/2)} \hat{\sigma}_{z}^{(N/2+N/5)} \rangle \sim N^{-2\Delta_z}$ 
has been analyzed on the AFM critical line $\thetaAFM(\alpha)$, as a function of $N$ and for values $\alpha \gtrsim 0.3$.
The coefficient $\Delta_z$ has been found to depart from the short-range value $\Delta_z =0.25$ for $\alpha \approx 2.25$, then 
decreasing and reaching $\Delta_z =0.2$ at $\alpha \approx 0.5$. 
The scaling coefficient $\Delta_z$ can be directly related to the scaling of the QFI: $\beta_c^+ = 1- 2 \Delta_z$. 
The results for $1- 2 \Delta_z$ found in Ref.~\cite{KoffelPRL2012} are reported as triangles in the inset of Fig.~\ref{fig5}(b) and compared to the values 
obtained in our numerical calculations (orange regions, corresponding to the data of the main panel).
They agree with our results except in the range $1 \lesssim \alpha \lesssim 2$ where 
they are systematically above our findings. 
As a check of our numerical calculations, the inset of Fig.~\ref{fig5}(b) shows the scaling $\beta(\alpha)$ obtained from 
the finite-size scaling of correlation functions 
$\langle\hat{\sigma}_{z}^{(N/2)} \hat{\sigma}_{z}^{(N/2+N/5)} \rangle \sim N^{\beta(\alpha)-1}$ (green dots)
and from the power-law decay
$\langle\hat{\sigma}_{z}^{(N/2)} \hat{\sigma}_{z}^{(N/2+r)} \rangle \sim r^{\beta(\alpha)-1}$ for $N=120$ (purple circles).
We see that the values of $\beta(\alpha)$ extracted in both cases are consistent with $\pow_c^+(\alpha)$ obtained via the analysis of the QFI. 
We thus conclude that the slight discrepancy within our numerical results and those of Ref.~\cite{KoffelPRL2012} is most likely due to 
the uncertainty in locating the critical point $\thetaAFM(\alpha)$.
It should be noticed however, that the interesting regime, where $\beta_c^+$ is notably different from the SR scaling, is found for values of
$\alpha \lesssim 0.5$, that were not analyzed in Ref.~\cite{KoffelPRL2012}.

The results reported in Fig.~\ref{fig5} strongly suggest the breakdown of conformal invariance along 
$\thetaAFM(\alpha)$ and $\thetaFM(\alpha)$ due to the LR coupling in (\ref{Hamiltonian}), 
at small-enough $\alpha$. 
The same breakdown has been previously inferred in \cite{VodolaNJP2016}, 
based on the scaling of the Von Neumann entropy. 
Oppositely to this quantity, the QFI density can be measured experimentally, 
yielding a direct way to probe the breakdown of conformal invariance in critical quantum systems. 
 
\begin{figure*}[!t]
\centering
\includegraphics[width=\textwidth]{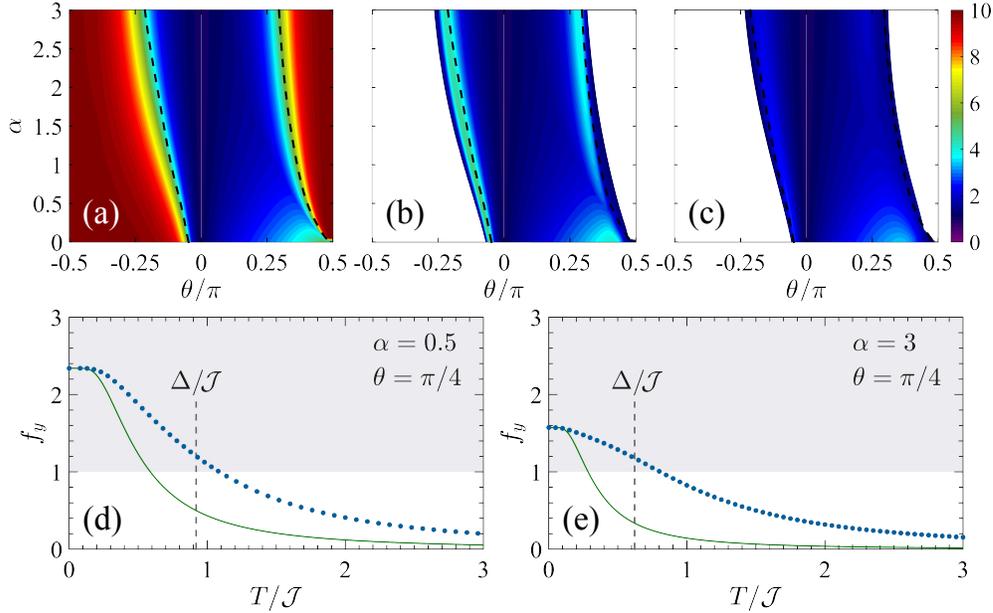} 
\caption{The upper panels show the QFI density $f_Q[\hat{\rho}_T, \hat{O}]$ as a function of $\theta$ and $\alpha$ 
for $T=0$ (a), $T/\Eunit=0.05$ (b) and $T/\Eunit=0.2$ (c). 
In the colored region the QFI density witnesses ME, since $f_Q>1$, 
while in the white regions $f_Q\leq1$. 
The operators chosen to calculate the QFI on the $\theta$-$\alpha$ plane are the same used at zero temperature in Section (\ref{multipartite}).
The black dashed lines signal the minima of the mass gap $\Delta$. 
The scarce appearance of the LRPM phase is a consequence of the very limited size of the chain
adopted here, $N = 10$.
The lower panels show the QFI density (dots) as a function of $T$,
compared with the analytical bound Eq.~(\ref{QFIscalT}) (solid line).
In panel (d), $\theta=\pi/4$ and $\alpha=0.5$ (LRPM region), 
in panel (e), $\theta=\pi/4$ and $\alpha=3$ (PM region).
The vertical dashed line is $T=\Delta$.
} 
\label{fig6}
\end{figure*}
 
\section{Multipartite entanglement at finite temperature} 
\label{finiteT}

The calculation of the QFI can be straightforwardly extended to finite-temperature states, using Eq.~(\ref{QFImixed}) and assuming thermal equilibrium 
$\hat{\rho}_T = e^{-\hat{H}/T} / {\rm Tr}[e^{-\hat{H}/T}]$, where 
$T$ is the temperature and the Boltzmann $k_B$ is set to $k_B=1$~\cite{HaukeNATPHYS2016, Gabbrielli2018, Frerot2018}. 
The QFI is obtained here by full numerical diagonalization of the Hamiltonian (\ref{Hamiltonian}) for fixed system sizes $N \leq  20$. 
The decay of the QFI density with $T$ characterizes the robustness of ME in the various phases.  
In Ref.~\cite{Gabbrielli2018} it has been shown that 
\be 
\frac{f_Q [\hat{\rho}_T, \hat{O} ]}{f_Q [\hat{\rho}_{T \to 0}, \hat{O}]} \geq \mathrm{tanh}^2 \Big(\frac{\Delta}{2T}\Big) \, \mu \, \frac{1+ e^{-\Delta/T}}{\mu + \nu \, e^{-\Delta/T}} \, ,
 \label{QFIscalT}
\ee
where $\mu$ and $\nu$ indicate the degeneracy of the ground state and the first excited state, respectively, and $\Delta$ is 
the mass gap, namely the finite difference between the ground-state energy and the energy of the first excited state in the thermodynamic limit.
Equation (\ref{QFIscalT}) identifies a temperature regime below a crossover temperature of the order of $\Delta$, 
where the QFI density is at least constant, lower bounded by its zero-temperature limit.

\begin{figure*}[!t]
\centering
\includegraphics[width=\textwidth]{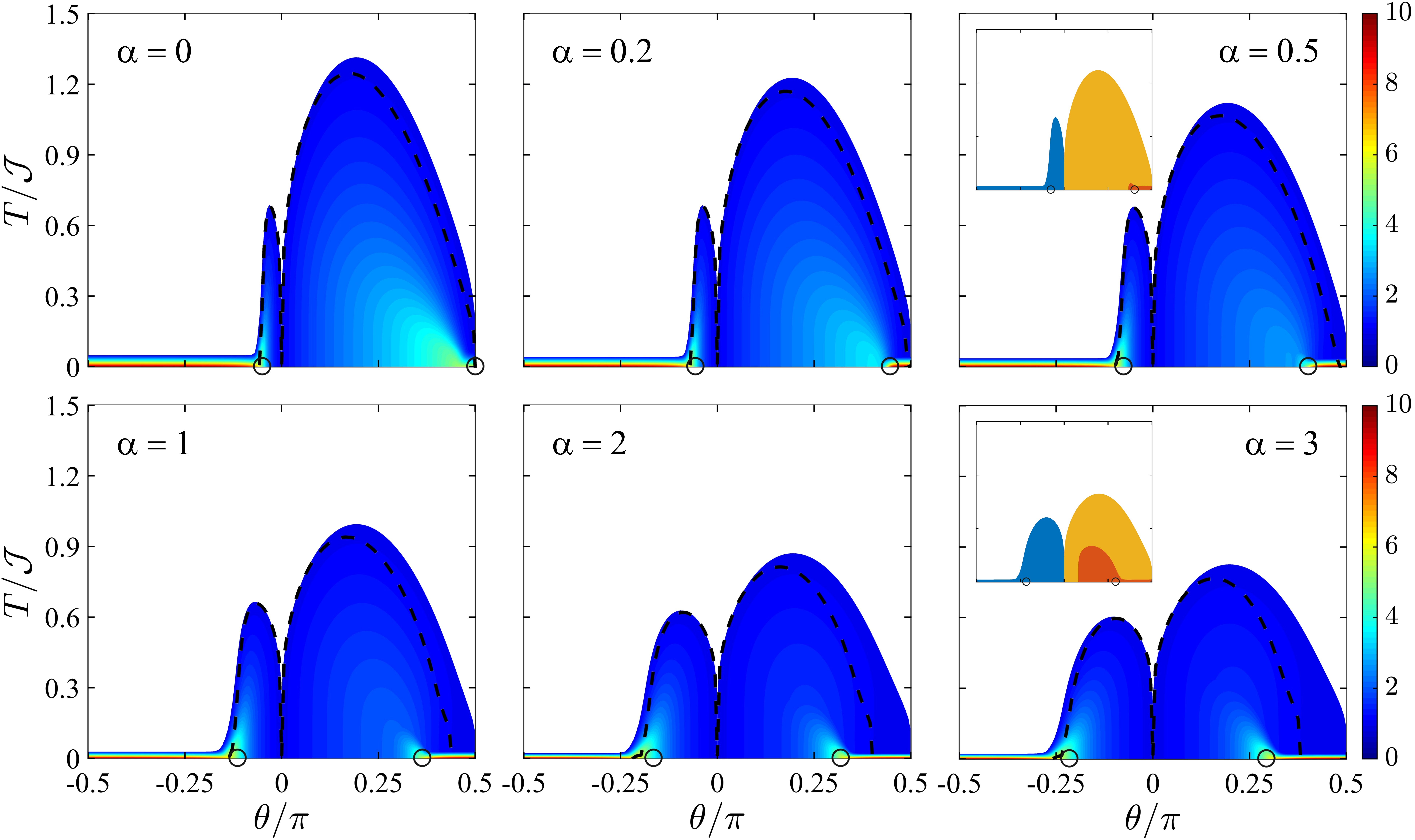} 
\caption{
QFI density $f_Q[\hat{\rho_T},\hat{O}]$ (color scale) of finite temperature states $\hat{\rho_T}$, on the $\theta$-$T$ plane.
Different panels correspond to different values of $\alpha$.
ME is witnessed in the colored regions.
A sudden decay from the high values at $T = 0$ is observed in the FM and AFM phases, due to the quasi double-degenerate ground state.
The QFI is optimized for any $\theta$, and $T$. As examples, the two insets in the panels for $\alpha = 0.5$ and $\alpha = 3$
display the optimal operators on the $\theta$-$T$ plane: blue for $\hat{J}_z$, yellow for $\hat{J}_y$ and red for $\hat{J}_z^{\rm(st)}$.
The black dashed line $\xi_{\rm R}^2=1$ encloses the area where the spin-squeezing parameter is able to detect entanglement. 
The black circles mark the position of the critical points.
In all the panels, $N=10$.
}
\label{fig7}
\end{figure*}

Figure~\ref{fig6}(a)-(c) shows the QFI density in the $\theta$-$\alpha$ phase diagram at different temperatures 
[color scale, where $T$ is expressed in unit of the magnetic coupling $\Eunit$ in Eq.~(\ref{Hamiltonian})], 
with the white regions corresponding to $f_Q \leq 1$.
In the FM and AFM ordered phases, the zero-temperature QFI is much larger than in the PM phase, see Fig.~\ref{fig6}(a).
Yet, this large value is lost abruptly for arbitrary small temperature $T$ (in the thermodynamic limit), 
reaching~\cite{Gabbrielli2018} 
\be \label{QFI0degerate}
f_Q[\hat{\rho}_{T \to 0}, \hat{O}]= \frac{2}{N} \Bigg( 
(\Delta \hat{O})^2_{\ket{\psi_{\rm gs}}} + (\Delta \hat{O})^2_{\ket{\psi_{\rm gs}'}} - 2 \big|\bra{\psi_{\rm gs}}\hat{O}\ket{\psi_{\rm gs}'}\big|^2 \Bigg),
\ee 
which is much lower than $f_Q[\ket{\psi_{\rm gs}}, \hat{O}]$. 
In Eq.~(\ref{QFI0degerate}), $\ket{\psi_{\rm gs}}$  and $\ket{\psi_{\rm gs}'}$ are 
the two quasi-degenerate ground states in the FM and AFM phases.
The discontinuity between $f_Q[\ket{\psi_{\rm gs}}, \hat{O}]$ and $f_Q[\hat{\rho}_{T \to 0}, \hat{O}]$ is due 
to the presence of a spontaneous symmetry breaking of the spin-flip $Z_2$ symmetry at $T=0$, 
resulting in a quasi-degeneracy of the ground state (that becomes an actual degeneracy in the thermodynamic limit only).
In Fig.~\ref{fig5}(b) we see that the QFI in the FM and AFM phases at finite temperature is not high enough to witness ME.
The QFI density remains high only close to the critical lines and, most interestingly, in the LRPM phase.
Indeed, in the LRPM phase
the ground state is nondegenerate also in the thermodynamic limit, as well as in the PM phase,
so that, according to Eq.~(\ref{QFIscalT}), the superextensive ME witnessed by the QFI at $T=0$, 
for $0 < \theta \leq \thetaAFM$ and $\alpha \leq 1$,
survives up to temperatures $T \approx \Delta$.
The typical decay of the QFI density in the LRPM and PM phases, compared to the lower bound Eq.~(\ref{QFIscalT}), is shown in Figs.~\ref{fig6}(d) and \ref{fig6}(e), respectively.

In Fig.~\ref{fig7} we plot the thermal phase diagram $\theta$-$T$ of the QFI density for different values of $\alpha$.
The colored region corresponds to $f_Q > 1$, where the QFI witnesses ME, while $f_Q \leq 1$ in the white region.  
We clearly distinguish two ``lobes'' on the FM and AFM sides of the phase diagram.
On the basis of the results above, we argue that the FM lobe (at $\theta<0$) disappears in the thermodynamic limit for $\alpha \leq 1$ due to the disappearance of the PM phase.
The boundary of the AFM lobe (at $\theta>0$) for small values of $\theta$ is well reproduced by the condition $\xi^2_{\rm R}=1$ (dashed line) corresponding to the thermal loss of spin squeezing.
Figure~\ref{fig7} also clearly shows the sharp decrease of the QFI in the FM and AFM phases at vanishingly small temperature $T/\Eunit\to 0$~\cite{Gabbrielli2018}.

We conclude this section noticing that, due to the absence of edge states, even in the open chains, the discussion about the phase stability against  temperature, performed by the scaling of the QFI density and based on  Eq. (\ref{QFIscalT}), does not suffer of deviations from edge contributions, similar to those hypothesized in \cite{quelle2016}. 

\section{Discussion and conclusions}

The scaling of the QFI calculated with respect to different collective operators yields
a characterization of the full phase diagram of the long-range Ising model that
can be probed in current quantum simulators, even for a limited ($N \lesssim 50$) number of spins.
This approach provides a clear signature of many physical effects that characterize the model, as the presence of a 
long-range paramagnetic phase at $\alpha \leq 1$, the change of the scaling along the critical massless lines for small values of $\alpha$ and a probe of the mean-field regime.
The long-range paramagnetic phase is particularly interesting. 
Here, ME can be also captured by the spin-squeezing parameter, that largely simplifies experimental detection and characterization of the state.  
The large ME (namely, superextensive QFI and inverse spin-squeezing parameter) found in the ground state is robust against temperature, being protected by an
energy gap that remains finite in the thermodynamic limit.
Furthermore, the LPRM phase can be addressed by preparing the ground state at $\theta=0$ and  
adiabatically increasing the coupling strength, without crossing any quantum phase transition.
Finally, we recall that the QFI studied here and the spin-squeezing parameter are directly related to metrological usefulness of a quantum state, see~\cite{PezzeRMP} for a recent review.
Therefore, we can conclude that the ground state of the long-range Ising model, especially in the robust long-range paramagnetic phase,
can be a resource for entanglement-enhanced metrology.

\ack{
We warmly thank Fabio Ortolani, Simone Paganelli, Augusto Smerzi, Luca Tagliacozzo, Andrea Trombettoni, and Davide Vodola for useful discussions and for the help with numerics.
The authors also acknowledge the participation to the workshop ``Entanglement in Quantum Systems'', 
held at the Galileo Galilei Institute for Theoretical Physics, Firenze, 21th May - 13th July 2018, where part of this work has been performed.
This work has been supported by the QuantEra project ``Q-Clocks'' and Empire project ``OQOQS''.\\

Short before the submission of this manuscript we became aware of a similar work done by M. Chiofalo and collaborators, to appear.

\appendix

\section{Numerical methods}
\label{AppNumerical}

The Ising Hamiltonian~(\ref{Hamiltonian}) is analytically treatable only in the special case of nearest-neighbor interaction ($\alpha=\infty$). 
In this case, exact results for the correlators appearing in Eq. (\ref{QFIcorr}) can be found for a close chain in the thermodynamic limit, see \cite{mussardo} and references therein.
Instead, when considering arbitrary interaction range $\alpha$, we must rely on numerical results. 
For short chains $N\leq20$ we performed an exact diagonalization to find the full spectrum and energy eigenstates,
from which a derivation of the energy gap, order parameter, fidelity susceptibility and QFI is possible. 
For $N>20$, we utilized an algorithm based on the density-matrix renormalization group~\cite{WhitePRL1992,WhitePRB1993}, an iterative variational technique optimized for the convergence of the ground state that provided us with all the spin-spin correlations, by which we could evaluate the QFI via the relation~(\ref{QFIcorr}), up to about $N\approx200$.

\section{Perturbative calculations}
\label{AppPerturb}

We have performed a perturbative calculation of the ground state for $\theta \to 0$.
At $\theta = 0$ the ground state is given by $\ket{\psi_{\mathrm{gs}}^{(0)}} = \spindownx^{\otimes N}$.
At first order in $\theta$, we find the normalized state
\be \label{IsingPerturbationState}
\ket{\psi_{\rm gs}^{(1)} (\theta)}\simeq \frac{1}{\sqrt{\pazocal{N}_\alpha}}  \, \Big(\ket{\psi_{\mathrm{gs}}^{(0)}} - \theta\,\mathcal{G}_N(\alpha)\,\ket{\psi_2^{(0)}} \Big) \, , 
\ee
where $\ket{\psi_2^{(0)}} = {\rm Sym}[\spindownx^{\otimes N-2} \spindownx^{\otimes 2}]$ is the unperturbed second excited state, given by the normalized
symmetric superposition of $N-2$ particles in $\spindownx$ and two particles in $\spinupx$, 
$\pazocal{N}_\alpha = 1 + \theta^2 \, \big[\mathcal{G}_N(\alpha)\big]^2$, 
\be
\mathcal{G}_N(\alpha) = \frac{N\,{\rm H}_{N,\,\alpha}-{\rm H}_{N,\,\alpha-1}}{\sqrt{8N(N-1)}}
\ee
and ${\rm H}_{N,\,\alpha}$ is the $N$-th generalized harmonic number of order $\alpha$.
In the limit $N\gg1$, the calculation of $\mathcal{G}_N(\alpha)$ involves handling hyperharmonic series, 
whose convergence is only attained for $\alpha>1$:
\be \label{IsingPerturbationFunction}
\mathcal{G}_N(\alpha) \approx 
\frac{1}{\sqrt{8}}
\times
\left\{
\begin{array}{ll} 
\zeta(\alpha) & {\rm for} \ \alpha>1 \\ 
\log N & {\rm for} \ \alpha=1 \\ 
\frac{1}{(1-\alpha)(2-\alpha)}\,N^{1-\alpha}   & {\rm for} \ 0\leq\alpha<1
\end{array}
\right.
\ee
where $\zeta(\alpha)$ is the Riemann zeta function.
The perturbative expansion in Eq.~(\ref{IsingPerturbationState}) is thus obtained
at fixed $|\theta|\ll1/\log{N}$ if $\alpha=1$ or $|\theta|\ll1/N^{1-\alpha}$ if $\alpha<1$, such to fulfill
the condition of validity of perturbation theory, $\,\theta\,\mathcal{G}_N(\alpha)\ll1$.
For $\alpha \leq 1$, this approximation breaks down in the thermodynamic limit. \\

{\it Ground-state energy and ferromagnetic critical line.} 
Using the nondegenerate perturbation theory it is also possible to evaluate the shift of the lowest energy levels due to the (small) interaction term. 
Up to the second order in $\theta$, we find 
$E_{\rm gs}^{(2)}/\Eunit = -N - 4\theta^2 [\mathcal{G}_N(\alpha)]^2$
and $E_{\rm ex}^{(2)}/\Eunit = 2-N + 2\theta\sqrt{8\frac{N-1}{N}}\mathcal{G}_N(\alpha) -12\frac{N-2}{N}\theta^2 [\mathcal{G}_N(\alpha)]^2$. 
Considering the form of $\mathcal{G}_N(\alpha)$ for $N\gg1$, at $\alpha>1$ we find $\Delta^{(1)} = 2 + 2\theta\zeta(\alpha)$ and $\Delta^{(2)} = 2 + 2\theta\zeta(\alpha) - \theta^2\zeta^2(\alpha)$.
The functional form of $\thetaFM(\alpha)$ is obtained, at first order, from $\Delta^{(1)}=0$, and, at second order, from $\Delta^{(2)}=0$.
Results are reported in the main text and in Fig.~\ref{fig1}. \\

{\it Quantum Fisher information.}
Using Eq.~(\ref{IsingPerturbationState}) it is possible to calculate the QFI for the different collective operators considered in the main text:
\be \label{IsingPerturbationFunction}
f_Q\big[\ket{\psi_n^{(1)}(\theta)},\hat{O}\big] = 
\left\{
\begin{array}{ll} 
\frac{8}{N}\,\theta^2\,\big[\mathcal{G}_N(\alpha)\big]^2 & {\rm for} \ \hat{O}=\hat{J}_x \\[3pt] 
1+\theta\sqrt{8\frac{N-1}{N}}\,\mathcal{G}_N(\alpha) & {\rm for} \ \hat{O}=\hat{J}_y \\[3pt]  
1-\theta\sqrt{8\frac{N-1}{N}}\,\mathcal{G}_N(\alpha) & {\rm for} \ \hat{O}=\hat{J}_z \\[3pt] 
\frac{8}{N}\,\theta^2\,\big[\mathcal{G}_N(\alpha)\big]^2 & {\rm for} \ \hat{O}=\hat{J}_x^{\rm(st)} \\[3pt]  
1-\theta\sqrt{\frac{8}{N(N-1)}}\,\mathcal{G}_N(\alpha) & {\rm for} \ \hat{O}=\hat{J}_y^{\rm(st)} \\[3pt]  
1+\theta\sqrt{\frac{8}{N(N-1)}}\,\mathcal{G}_N(\alpha) & {\rm for} \ \hat{O}=\hat{J}_z^{\rm(st)}
\end{array}
\right.
\ee
These predictions are in agreement with the behavior found numerically, see Fig.~\ref{fig4}.  \\

{\it Fidelity susceptibility.}
Let us now evaluate, at fixed $N$ and $\theta$ (such that $\theta\,\mathcal{G}_N(\alpha)\ll1$) 
the fidelity between the ground states corresponding to
two close interaction ranges $\alpha$ and $\alpha+\delta\alpha$: 
at leading order in $\theta$,
\be 
\pazocal{F}_{\alpha,\,\alpha+\delta\alpha} = \Big| \big\langle\psi_{\rm gs}^{(1)}(\theta,\alpha) \big|
\psi_{\rm gs}^{(1)}(\theta,\alpha+\delta\alpha)\big\rangle \Big|^2 \approx 1 - \frac{\theta^2}{2} 
\bigg[\mathcal{G}_N(\alpha+\delta\alpha)-\mathcal{G}_N(\alpha)\bigg]^2.
\ee
This yields the fidelity susceptibility~\cite{zanardi2007}
\be
\chi_\alpha = 
-2 \lim_{\delta\alpha\to0}\frac{\log\pazocal{F}_{\alpha,\,\alpha+\delta\alpha}}{(\delta\alpha)^2}  = \theta^2 \left[\frac{\partial}{\partial\alpha}\mathcal{G}_N(\alpha)\right]^2  \, .
\ee
We find that $\chi_\alpha$ asymptotically scales as $N^{2(1-\alpha)} \,(\ln{N})^2$ for large $N$ when $\alpha<1$, 
whereas it saturates to a constant $\pazocal{O}(1)$ when $\alpha>1$.

\section{Variational calculation}
\label{AppVariational}

For $\alpha=0$ we use a variational ansatz to calculate the ground state and the QFI.
We rewrite Eq.~(\ref{Hamiltonian}) as
\be \label{HamSym}
\frac{\hat H} {2 \Eunit} =  \hat{J}_z^2 \sin \theta + \hat{J}_x \cos \theta.
\ee
It is well known that the model (\ref{HamSym}) can be studied by restricting to a basis of eigenstates $\{ \ket{\mu} \}$ of the collective operator $\hat{J}_z$
($\hat{J}_z \ket{\mu} = \mu \ket{\mu}$, with $\mu = -N/2, -N/2+1, ..., N/2$) made of $N+1$ orthogonal states symmetric under particle exchange.
We search the ground state making use of the Gaussian variational ansatz
\be \label{statevar}
\ket{\psi_{\rm gs}} = \sum_{\mu = -N/2}^{N/2} \frac{e^{-\mu^2/(4 \sigma^2)}}{(2 \pi \sigma^2)^{1/4}} \ket{\mu}, 
\ee
where the width $\sigma$ is the sole variational parameter. 
We further assume $N \gg1$ and a sufficiently localized wavepacket so to neglect border effects. 
We minimize the energy $E_{\rm gs} = \bra{\psi_{\rm gs}} \hat{H} \ket{\psi_{\rm gs}}$, 
\be \label{Evar}
\frac{E_{\rm gs}}{2 \Eunit} \approx  \sigma^2 \sin \theta -  \frac{N}{2} \cos \theta  \bigg[ 1 - \frac{2}{N^2} \Big( \sigma^2-\frac{1}{4} \Big)\bigg] e^{-\frac{1}{8 \sigma^2}}, 
\ee
where we have used $\sqrt{(N/2)(N/2+1)} \approx N/2$ and taken the continuous limit for $\mu$.
Within the same approximations, the QFI calculates as $f[\ket{\psi_{\rm gs}}, \hat{J}_y] = N/(4 \sigma^2)$.
The equation $\frac{d E_{\rm gs}}{d \sigma^2}=0$ gives 
\be \label{EqVar}
e^{-\frac{1}{8 \sigma^2}} \Big( \frac{N}{16 \sigma^4} - \frac{1}{8 N \sigma^2} - \frac{1}{N} \Big) \cos \theta = \sin \theta.
\ee
For $\sigma^2 \gg 1$ we can neglect the term $e^{-\frac{1}{8 \sigma^2}}$ in Eq.~(\ref{EqVar})
and we obtain
\be \label{SigmaVar}
\sigma^2 = \frac{N}{4} \frac{1}{\sqrt{1+N\tan \theta}}, \quad {\rm and} \quad f[\ket{\psi_{\rm gs}}, \hat{J}_y] = \sqrt{1 + N \tan \theta},
\ee
recovering $\sigma^2 = N/4$ and $f[\ket{\psi_{\rm gs}}, \hat{J}_y]=1$, respectively, at $\theta=0$.
We can distinguish different behaviors and limits.
For $0 < \theta  \ll 1/N$ (corresponding to the so-called Rabi regime for the Josephson junction~\cite{PezzePRA2005, PezzeRMP}), 
we obtain $\sigma^2 = \frac{N}{4}(1 - \frac{N \theta}{2})$ and $f[\ket{\psi_{\rm gs}}, \hat{J}_y] = 1 + \frac{N \theta}{2}$,
which is exactly the perturbative prediction reported in Eq.~(\ref{IsingPerturbationFunction}) for $\alpha=0$.
For $1/N \ll \tan \theta  \ll N$ (corresponding to the so-called Josephson regime~\cite{PezzePRA2005, PezzeRMP}), we obtain
$\sigma^2  = \sqrt{\frac{N}{16 \tan \theta}}$ and thus $f[\ket{\psi_{\rm gs}}, \hat{J}_y] = \sqrt{N\tan \theta}$,
in agreement with our numerical calculations predicting a scaling $N^{1/2}$ of the QFI density, see Fig.~\ref{fig2}.
For $\tan \theta \gg N$ (Fock regime~\cite{PezzePRA2005, PezzeRMP}), Eq.~(\ref{SigmaVar}) predicts $\sigma=0$, corresponding to the symmetric 
Dicke state limit of Eq.~(\ref{statevar}), $\ket{\psi_{\rm gs}} = \ket{\mu=0}$, and for this state the QFI is equal to $f[\ket{\psi_{\rm gs}}, \hat{J}_y] = N/2+1$.
We can thus roughly locate a diverging derivative of the QFI (signaling the critical point) at $\theta = \arctan N$, that is $\theta \to \pi/2$ for $N \to \infty$.


\begin{thebibliography}{100}
\bibitem{notaME}
Multipartite entanglement is quantified by the number of particles in the largest nonseparable subset~\cite{GuhnePR2009, PezzeRMP}.
A pure state of $N$ particles is $k$-separable (also indicated as $k$-producible in the literature) 
if it can be written as $\ket{\psi_{k\textrm{-}{\rm sep}}} = \ket{\psi_{N_1}} \otimes \ket{\psi_{N_2}} \otimes ... \otimes  \ket{\psi_{N_M}}$, 
where $\ket{\psi_{N_l}}$ is a state of $N_l \leq k$ particles that does not factorize and $\sum_{l=1}^M N_l = N$.
A mixed state is $k$-separable if it can be written as a mixture of $k$-separable pure states.
A state that is $k$-separable but not $(k-1)$-separable is called $k$-particle entangled: it contains at least one state of $k$ particles that does not factorize.
\bibitem{notaLL}
For comparison, the fits up to $N=50$ provide the same values of the fitting parameters for the panels (a)-(c), within a $10^{-2}$ precision. 
On the contrary, for panel (d) we obtain $f_y = 1 + 0.12 \, N^{0.62}$.
\end{thebibliography}
\end{document}